\documentclass[letterpaper, oneside, 12pt]{article}
\usepackage{natbib}
\usepackage{amsmath}
\usepackage{amssymb}
\usepackage{makeidx}
\usepackage{graphicx}
\usepackage{amsfonts}
\usepackage{latexsym}
\usepackage{subfigure}
\usepackage{url}
\usepackage{algpseudocode}
\usepackage{longtable}
\usepackage{multirow}
\usepackage{fancyvrb}
\usepackage{float}
\usepackage{cancel}
\usepackage[hmargin=1.0in,vmargin=1.0in]{geometry}
\usepackage{setspace}
\usepackage{rotating}
\usepackage{enumerate}
\usepackage{comment}
\usepackage{booktabs}
\usepackage{color}
\usepackage{algorithm}
\usepackage{authblk}
\newcommand\Algphase[1]{%
\vspace*{-.7\baselineskip}\Statex\hspace*{\dimexpr-\algorithmicindent-2pt\relax}\rule{\textwidth}{0.4pt}%
\Statex\hspace*{-\algorithmicindent}\textbf{#1}%
\vspace*{-.7\baselineskip}\Statex\hspace*{\dimexpr-\algorithmicindent-2pt\relax}\rule{\textwidth}{0.4pt}%
}
\usepackage{setspace}
\DefineVerbatimEnvironment{Code}{Verbatim}{fontsize=\small}
\newcommand{\ignore}[1]{}
\newcommand{\bmu}{ \mbox{\boldmath $\mu$} }
\newcommand{\bSig}{ \mbox{\boldmath $\Sigma$} }
\newcommand{\bbeta}{ \mbox{\boldmath $ \beta $} }

\newcommand{\calK}{{\cal K}}

\newcommand{\tSig}{\widetilde \bSig}

\newcommand{\tC}{\widetilde \bC}
\newcommand{\tM}{\widetilde \bM}

\newcommand{\bLam}{ \mbox{\boldmath $\Lambda$} }

\newcommand{\taus}{\tau^2}
\newcommand{\sigs}{\sigma^2}
\newcommand{\given}{\,|\,}
\newcommand{\btheta}{ \mbox{\boldmath $ \theta $} }
\newcommand{\boeta}{ \mbox{\boldmath $ \eta $} }
\newcommand{\bOmega}{ \mbox{\boldmath $ \Omega $} }
\newcommand{\bzero}{\textbf{0}}

\newcommand{\T}{\top}

\newcommand{\bA}{\textbf{A}}
\newcommand{\bb}{\textbf{b}}
\newcommand{\bB}{\textbf{B}}
\newcommand{\bc}{\textbf{c}}
\newcommand{\bC}{\textbf{C}}

\newcommand{\bD}{\textbf{D}}

\newcommand{\bF}{\textbf{F}}
\newcommand{\bff}{\textbf{f}}

\newcommand{\bI}{\textbf{I}}
\newcommand{\bg}{\textbf{g}}
\newcommand{\bG}{\textbf{G}}

\newcommand{\bL}{\textbf{L}}

\newcommand{\bM}{\textbf{M}}

\newcommand{\bP}{\textbf{P}}
\newcommand{\bs}{\textbf{s}}
\newcommand{\bS}{\textbf{S}}

\newcommand{\bu}{\textbf{u}}

\newcommand{\bv}{\textbf{v}}
\newcommand{\bV}{\textbf{V}}
\newcommand{\bw}{\textbf{w}}

\newcommand{\bx}{\textbf{x}}
\newcommand{\bX}{\textbf{X}}
\newcommand{\by}{\textbf{y}}

\newcommand{\br}{\textbf{r}}

\newcommand{\bz}{\textbf{z}}
\newcommand{\tildebC}{\widetilde{\bC}}

\usepackage[compact]{titlesec}
\titlespacing{\section}{0pt}{*0}{*0}
\titlespacing{\subsection}{0pt}{*0}{*0}
\titlespacing{\subsubsection}{0pt}{*0}{*0}

\usepackage{etoolbox}

\makeatletter
\patchcmd{\ttlh@hang}{\parindent\z@}{\parindent\z@\leavevmode}{}{}
\patchcmd{\ttlh@hang}{\noindent}{}{}{}
\makeatother

\makeatletter
\newenvironment{breakablealgorithm}
  {% \begin{breakablealgorithm}
   \begin{center}
     \refstepcounter{algorithm}% New algorithm
     \hrule height.8pt depth0pt \kern2pt% \@fs@pre for \@fs@ruled
     \renewcommand{\caption}[2][\relax]{% Make a new \caption
       {\raggedright\textbf{\ALG@name~\thealgorithm} ##2\par}%
       \ifx\relax##1\relax % #1 is \relax
         \addcontentsline{loa}{algorithm}{\protect\numberline{\thealgorithm}##2}%
       \else % #1 is not \relax
         \addcontentsline{loa}{algorithm}{\protect\numberline{\thealgorithm}##1}%
       \fi
       \kern-2pt\kern-2pt
     }
  }{% \end{breakablealgorithm}
     \kern-2pt\hrule\relax% \@fs@post for \@fs@ruled
   \end{center}
  }
\makeatother

\usepackage{xr}
\begin{document}

\setlength{\abovedisplayskip}{0pt}
\setlength{\belowdisplayskip}{0pt}

\title{Efficient algorithms for Bayesian Nearest Neighbor Gaussian Processes}

\author[1]{Andrew O. Finley}
\author[2]{Abhirup Datta}
\author[3]{Bruce C. Cook}
\author[3]{Douglas C. Morton}
\author[4]{Hans E. Andersen}
\author[5]{Sudipto Banerjee}
\affil[1]{Michigan State University}
\affil[2]{Johns Hopkins University}
\affil[3]{National Aeronautics and Space Administration}
\affil[4]{United States Forest Service}
\affil[5]{University of California, Los Angeles}

\date{\today}
\maketitle

\begin{abstract}
We consider alternate formulations of recently proposed hierarchical Nearest Neighbor Gaussian Process (NNGP) models \citep[][]{nngp} for improved convergence, faster computing time, and more robust and reproducible Bayesian inference. Algorithms are defined that improve CPU memory management and exploit existing high-performance numerical linear algebra libraries. Computational and inferential benefits are assessed for alternate NNGP specifications using simulated datasets and remotely sensed light detection and ranging (LiDAR) data collected over the US Forest Service Tanana Inventory Unit (TIU) in a remote portion of Interior Alaska. The resulting data product is the first statistically robust map of forest canopy for the TIU.
\end{abstract}

\section{Introduction}\label{sec:intro}

\noindent As spatial statisticians confront massive datasets with locations $\sim$$10^6$ and increasingly demanding inferential questions, several existing approaches that once seemed attractive for locations in the order of $10^4$ become impractical. Recent methodological developments within the burgeoning literature on this subject aim to deliver massively scalable spatial processes. \cite{sunligenton11} and \cite{banerjee2017} provide background and more current work (also see references therein), respectively, in this area. A recent contribution by \cite{heaton2017} is particularly useful as it provides an overview of modeling approaches for large spatial data that are under active development, and a comparison of these approaches based on the analysis of a common dataset in the form of a ``friendly competition.'' In addition to Nearest Neighbor Gaussian Process \citep[NNGP:][]{nngp} models, the comparison presented by \cite{heaton2017} considered reduced rank predictive processes \citep{ban08, finley2009}, covariance tapering \citep{Furr:Sain:10, spam}, gapfilling \citep{gapfill}, metakriging \citep{GuhBan16}, spatial partitioning \citep{sang2011covariance, barbian2017spatial}, fixed rank kriging \citep{Cressie_2008,zammit2017frk}, multiresolution approximation \citep{katzfussmultires}, stochastic partial differential equations \citep{rinla}, lattice kriging \citep{nychka2015multiresolution}, and local approximate Gaussian processes \citep{gram14, gramacy2016laGP}. The comparison was based on out-of-sampled predictive performance and, to a lesser extent, computing time for a moderately sized simulated and real dataset comprising 105,569 observations. Comparisons showed NNGP models yielded highly competitive predictive performance and computation time. 

%Current demands for large scale spatial analysis, such as the scientific application describe below, require statistical inference on underlying spatial processes or random fields at arbitrary resolutions. Bayesian inference is attractive here as it supplies full posterior predictive distributions for the outcomes and the latent processes at arbitrary locations. 
With a few exceptions, e.g., \cite{Furr:Sain:10} and \cite{gramacy2016laGP}, the literature on scalable spatial process models has focused primarily on theoretical and methodological developments with little attention to the algorithmic details needed for effectively applying them. For example, \cite{nngp} implement a ``sequential'' Gibbs sampler that involves updating a high-dimensional latent random effect vector and is prone to high autocorrelations and slow convergence. Most of the aforementioned articles do not discuss %strategies possible reparametrizations for massively scalable processes and 
how researchers can, in practice, exploit high-performance computing libraries to obviate expensive numerical linear algebra (e.g., expensive matrix multiplications and factorizations) and deliver full Bayesian inference for massive spatial datasets. We address this gap for the NNGP models here by outlining three alternate formulations that are significantly more efficient for practical implementation than \cite{nngp}. Along with the accompanying code supplied with this manuscript, our intended contribution is well aligned with recent emphasis on reproducible analysis for challenging data analysis in the context of massive spatial datasets.               

%Forests cover approximately one third of the land surface and provide a broad range of critical ecosystem services \citep{Sexton16}.  Forest characteristics vary by biome, climate, soils, and topography \citep{Holdridge67}, with fine-scale variability in forest structure and composition from disturbance processes such as wind, fire, insects, and human activity (e.g., logging, fuel wood collection, and forest fragmentation). Forests store large amounts of carbon in aboveground biomass and organic soils \citep[e.g.,][]{Turetsky11}, and evidence suggests they play a critical role in the global land carbon sink \citep{pan11, Schimel11}.  Ongoing efforts to reduce forest carbon emissions and enhance carbon uptake (e.g., to Reducing Emissions from Deforestation and forest Degradation, REDD+), provide strong motivation for regional assessments of contemporary forest carbon stocks, {\color{blue} with uncertainty reporting}, in support of climate mitigation efforts, forest management, and carbon cycle science.

Our motivating scientific application concerns forest resource monitoring efforts and, in particular, to create fine resolution canopy height predictions using remotely sensed data collected at over 5 million locations. Spatially explicit estimates of forest canopy height are key inputs to a variety of ecosystem and Earth system modeling efforts \citep{finney04, Hurtt04, stratton06, lefsky10, klein15}. These and similar applications seek inference about forest canopy height model parameters and predictions that can be propagated through the subsequent computer models of ecosystem function to yield more robust error quantification. Bayesian inference is attractive here as it supplies full posterior predictive distributions for the outcomes and for the latent process at arbitrary locations in the region of interest. 

The remainder of this article proceeds as follows. Section~\ref{sec:nngp} provides a brief overview of NNGP models and their computational aspects. This is followed by three distinct and efficient alternate formulations: the collapsed NNGP model, an NNGP model for the outcomes themselves (with no latent process), and a conjugate NNGP model that allows MCMC-free inference. Section~\ref{sec:sim} offers several detailed simulation experiments on model performance and assessment and also presents a detailed analysis of the US Forest Service Tanana Inventory Unit (TIU) dataset. Finally, Section~\ref{sec:summary} concludes the manuscript with a summary and an eye toward future work.   

\section{Nearest Neighbor Gaussian Processes }\label{sec:nngp}
%Spatial regression models are widely used to analyze geo-indexed datasets in paradigms where available predictors do not adequately capture the spatial variability of the response. Under such settings, a random effect is often introduced to account for any residual spatial trends not explained by the predictors. 
\noindent Let $y(\bs_i)$ and $\bx(\bs_i)$ denote the response and the predictors observed at location $\bs_i$, $i=1,2,\ldots,n$. A spatial linear mixed model posits $y(\bs_i)= \bx(\bs_i)^\top\bbeta + w(\bs_i) + \epsilon(\bs_i)$, where the random effect $w(\bs_i)$ sums up the effect of unknown or unobserved spatial covariates, and $\epsilon(\bs_i)$ denotes the independent and identically observed noise. Gaussian Processes (GP) are commonly used for modeling the unknown surface $w(\bs)$. In particular, $w(\bs) \sim GP(0, C(\cdot,\cdot \given \btheta))$ implies that $\bw=(w(\bs_1),w(\bs_2),\ldots,w(\bs_n))^\top$ is Gaussian with mean zero and covariance $\bC=(c_{ij}) $, where $c_{ij}=C(\bs_i, \bs_j \given \btheta)$ and $\btheta$ denotes the GP covariance parameters. A popular choice for $C(\cdot,\cdot \given \btheta)$ is the Mat\'ern covariance function specified as:
\begin{equation}\label{eq:matern}
C(\bs_i, \bs_j; \sigs,\phi,\nu) = \frac{\sigs}{2^{\nu - 1}\Gamma(\nu)}(||\bs_i-\bs_j||\phi)^{\nu}\calK_{\nu}(||\bs_i-\bs_j||\phi);\; \phi > 0, \nu >0,
\end{equation}
where $\btheta=\{\sigs,\phi,\nu\}$ and $\calK$ denotes the Bessel function of second kind. Customary Bayesian hierarchical models are constructed as 
\begin{equation}\label{eq:likewy}
p(\bbeta,\btheta,\taus) \times N(\bw \given \bzero, \bC) \times N(\by \given \bX\bbeta + \bw, \taus \bI) \;,
\end{equation}
where $p(\bbeta,\btheta,\taus)$ is specified by assigning priors to $\bbeta$, $\btheta$ and $\taus$. %The hierarchical setup in (\ref{eq:likewy}) offers a rich and flexible framework for modeling spatial data and allows for inference on the spatial random effects $\bw$. This is often of scientific interest for researchers, as the residual spatial surface $w(\bs)$ may offer insight on local trends or the effect of unobserved covariates. However, 
When $n$ is very large, implementing ($\ref{eq:likewy}$) poses multiple computational roadblocks. Firstly, storing the matrix $\bC$ requires $O(n^2)$ dynamic memory. Furthermore, evaluating $N(\bw\given \bzero,\bC)$ involves factorizations (e.g., Cholesky) that require $O(n^3)$ floating point operations (flops) to solve linear systems involving $\bC$ and computing $\det(\bC)$. Finally, predicting the response at $K$ new locations require an additional $O(Kn^2)$ flops. Alternative parametrizations such as integrating $\bw$ out of (\ref{eq:likewy}) shrinks the size of the parameter space, but does not obviate these computational bottlenecks. Even for moderately large spatial datasets, say with with $\sim 10^4\mbox{--}10^5$ locations, these memory and storage demands become prohibitive. For the TIU dataset with $5 \times 10^6$ locations, implementing (\ref{eq:likewy}) is practically impossible. 

%In recent years, the ubiquity of such massive geo-referenced datasets in the fields of forestry, ecology, and climate sciences, has resulted in a deluge of statistical methods seeking scalable solutions for large spatial data analysis. Popular approaches include low rank models, composite likelihoods, nearest neighbor approximations, covariance tapering, among many others. 
%The recent expository article by \cite{wiresnngp} contains a substantial review of methods for spatial big data. The burgeoning literature on this subject is already too vast to be summarized here and we refer the reader to the Section 2 of \cite{wiresnngp} %of the recent expository article by   for an overview of extant approaches. 
%\textbf{AD/AF: Can we get a broader reference here than the WIREs paper?} 
%Very recently \cite{nngp} proposed Nearest Neighbor Gaussian Processes (NNGP) for large spatial datasets, A large subclass of these approaches achieve scalability via introduction of sparsity into the precision structure of the random effects $\bw$. 
As mentioned in the Introduction, we pursue massive scalability for full Bayesian inference exploiting the Nearest Neighbor Gaussian Processes (NNGP). The underlying idea is familiar 
%NNGP is a well defined Gaussian Process whose finite dimensional realizations follow multivariate Gaussian distribution with sparse precision matrices.
%Sparsity is ubiquituous 
in graphical models \citep[see, e.g.,][]{lauritzen96,murphy2012}. 
The joint distribution for a random vector $\bw$ can be looked upon as a directed acyclic graph (DAG). 
%We write $p(\bw) = p(w_1)\prod_{i=2}^n p(w_i \given w_1,\ldots, w_{i-1})$ where $w_i = w(\bs_i)$. The DAG corresponding to this factorization is shown in Figure~\ref{fig: full_graph} for $n=7$. 
% %This is specific to an ordering of the nodes, which is referred to as a topological ordering of the graph. 
% \begin{figure}[h]
% 	\begin{center}
% 		\subfigure[Full graph]{\includegraphics[width=2.5in,height=2.5in]{nnfull.png}\label{fig: full_graph}}
% 		\subfigure[Sparse graph]{\includegraphics[width=2.5in,height=2.5in]{nn3.png}\label{fig: sparse_graph}}
% 	\end{center}
% 	\caption{Sparsity using directed acyclic graphs} \label{fig: graph}
% \end{figure}
% Figure~\ref{fig: sparse_graph} shows the DAG when some of the edges are deleted so as to retain at most $3$ ``nearest neighbors'' in the conditional probabilities. The resulting joint density is
% \begin{align*}
% & p(w_1)\times p(w_2\given w_1)\times p(w_3\given w_1,w_2)\times p(w_4\given w_1,w_2,w_3) \\
% &\qquad \times p(w_5\given \cancel{w_1}, w_2,w_3,w_4) \times p(w_6\given w_1, \cancel{w_2},\cancel{w_3}, w_4,w_5) \times p(w_7\given w_1,w_2,\cancel{w_3},\cancel{w_4},\cancel{w_5}, w_6)\; . 
% \end{align*}
We write $p(\bw) = p(w_1)\prod_{i=2}^n p(w_i \given \mbox{Pa}[i])$, where $w_i \equiv w(\bs_i)$ and $\mbox{Pa}[i] = \{w_1,w_2,\ldots,w_{i-1}\}$ is the set of parents of $w_i$. We can construct sparse models for $\bw$ by shrinking the size of $\mbox{Pa}[i]$. In spatial contexts, this can be done by defining $\mbox{Pa}[i]$ to be the set of $w(\bs_j)$'s corresponding to a small number $m$ of nearest neighboring locations of $\bs_i$. Approximations resulting from such shrinkage have been originally proposed by \cite{ve88} and studied and exploited by \cite{stein04,stroud14,nngp,dnngp,Huang16}. %\textbf{AD: Supply references to Vecchia, Stein et al, Stroud et al and Gramacy's work.} 
The NNGP builds upon previous ideas and extends finite-dimensional likelihood approximations to well-defined sparsity-inducing Gaussian processes for estimating (\ref{eq:likewy}). 

Working with multivariate Gaussian densities makes the connection between conditional independence in DAGs and sparsity abundantly clear. %Consider an $n\times 1$ random vector $w = (w_1,w_2,\ldots,w_n)^{\T}$ distributed as $N(0,\bC)$. 
We can write the multivariate Gaussian density $N(\bw\given \bzero, \bC)$ as %$p(w_1)\prod_{i=1}^{n}p(w_i\given w_1, w_2, w_{i-1}))$. This can also be written as 
a linear model,
\begin{align*}
 w_1 &= 0 + \eta_1\; \quad \mbox{ and }\; \quad w_i = a_{i1}w_1 + a_{i2}w_2 +  \cdots + a_{i,i-1}w_{i-1} + \eta_i\; \mbox{ for } i=2,\ldots,n\; ,
% \Longrightarrow w &= Aw + \eta;\quad \eta \sim N(0, D)
\end{align*}
or, more compactly, simply as $\bw = \bA\bw + \boeta$, where $\bA$ is $n\times n$ strictly lower-triangular with elements $a_{ij} = 0$ whenever $j \geq i$ and $\boeta \sim N(\bzero, \bD)$ and $\bD$ is diagonal with entries $d_{11} = \mbox{var}(w_1)$ and $d_{ii} = \mbox{Var}(w_i\given \{w_j : j < i\})$ for $i=2,\ldots,n$. 

From the structure of $\bA$ it is evident that $\bI-\bA$ is nonsingular and $\bC = (\bI-\bA)^{-1}\bD(\bI-\bA)^{-\top}$. %The possibly nonzero elements of $\bA$ and $\bD$ are completely determined by the matrix $\bC$. 
For any matrix $\bM$ and set of indices $I_1$, $I_2 \subseteq \{1,2,\ldots,n\}$, let $\texttt{\bM}[I_1,I_2] $ denote the submatrix of $\bM$ formed by the rows indexed by $I_1$ and columns indexed by $I_2$.  %the the row-indices  $\texttt{\bA[i,j]}$, $\texttt{\bD[i,j]}$ and $\texttt{\bC[i,j]}$ denote the $(i,j)$-th entries of $\bA$, $\bD$ and $\bC$, respectively. 
Note that $\texttt{\bD[1,1] = C[1,1]}$ and the first row of $\bA$ is $\bzero$. A pseudocode to compute the remaining elements of $\bA$ and $\bD$ is
\begin{align}\label{eq: pseudocode_full_gaussian}
\begin{array}{ll}
&\texttt{for(i in 1:(n-1))} \texttt{ \{ } \\ 
&\qquad\texttt{\bA[i+1,1:i] = solve(\bC[1:i,1:i], \bC[1:i,i+1])} \\
&\qquad\texttt{\bD[i+1,i+1] = \bC[i+1,i+1] - dot(\bC[i+1,1:i],\bA[i+1,1:i])}  \\
& \texttt{ \},}
\end{array}
\end{align}
where  %$\texttt{\bA[i+1,1:i]}$ is the $1\times \texttt{i}$ row vector comprising the possibly nonzero elements of the $\texttt{i+1}$-th row of $A$, $\texttt{\bC[1:i,1:i]}$ is the $\texttt{i}\times \texttt{i}$ leading principal submatrix of $\bC$, $\texttt{\bC[1:i, i]}$ is the $\texttt{i}\times 1$ row vector formed by the first $\texttt{i}$ elements in the $\texttt{i}$-th column of $\bC$,  $\texttt{\bC[i, 1:i]}$ is the $1\times \texttt{i}$ row vector formed by the first $\texttt{i}$ elements in the $\texttt{i}$-th row of $\bC$, 
$\texttt{1:i}$ denotes the set $\{1,2,\ldots,i\}$, $\texttt{solve(B,b)}$ computes the solution $\bx$ for the linear system $\bB\bx = \bb$, and $\texttt{dot(u,v)}$ denotes the inner-product between two vectors $\bu$ and $\bv$.

\begin{figure}[h]
	\begin{center}
		\subfigure[$\bI-\bA$]{\includegraphics[width=4.5cm,trim={2cm 2cm 0cm 0cm},clip]{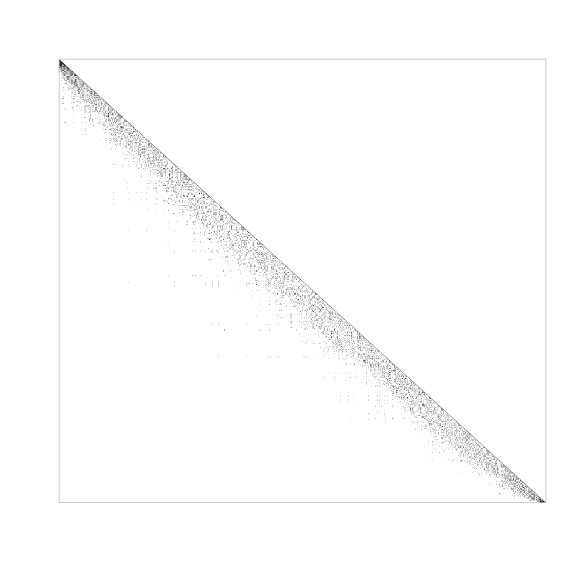}\label{fig: I-A}}
		\subfigure[$\bD^{-1}$]{\includegraphics[width=4.5cm,trim={2cm 2cm 0cm 0cm},clip]{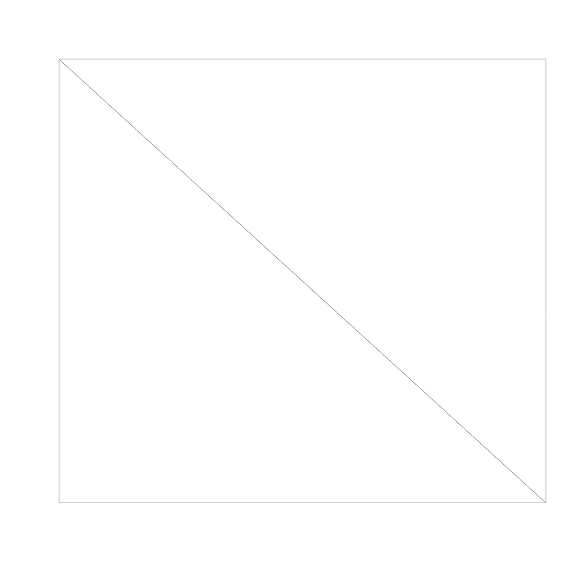}\label{fig: D.inv}}
		\subfigure[$\tilde\bC^{-1}$]{\includegraphics[width=4.5cm,trim={2cm 2cm 0cm 0cm},clip]{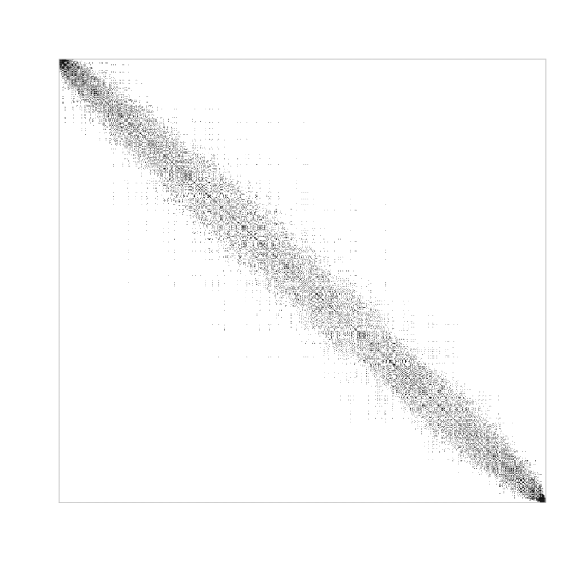}\label{fig: K.theta.inv}}
	\end{center}
	\caption{Structure of the factors making up the sparse $\tildebC^{-1}$ matrix.} \label{fig: NNGP_Chol}
\end{figure}

The above pseudocode computes the Cholesky decomposition of $\bC$. % as $\bC = \bL\bD\bL^{\top}$ is the Cholesky decomposition, then $\bL = (\bI-\bA)^{-1}$. 
There is, however, no apparent gain to be had from the preceding computations since, as the loop runs into higher values of $\texttt{i}$ closer to $\texttt{n}$, the dimension of $\texttt{\bC[1:i,1:i]}$ increases. Consequently, one will need to solve larger and larger linear systems and the computational complexity remains $O (n^3)$. Nevertheless, it immediately shows how to exploit sparsity if we set some elements in the lower triangular part of $\bA$ to be zero. For example, suppose we permit no more than $\texttt{m}$ elements in each row of $\bA$ to be nonzero. Let $\texttt{N[i]}$ be the set of indices $\texttt{j} < \texttt{i}$ such that $\texttt{\bA[i,j]} \neq 0$. One can then compute the elements of $\bA$ and $\bD$ as:
\begin{align}\label{eq: pseudocode_sparse_gaussian}
\begin{array}{ll}
&\texttt{for(i in 1:(n-1)} \texttt{ \{ } \\ 
&\qquad\texttt{\bA[i+1,N[i+1]] = solve(\bC[N[i+1],N[i+1]], \bC[N[i+1],i+1])} \\
&\qquad\texttt{\bD[i+1,i+1] = \bC[i+1,i+1] - dot(\bC[i+1, N[i+1]], \bA[i+1,N[i+1]])}\\
& \texttt{\}.}
\end{array}
\end{align}
In (\ref{eq: pseudocode_sparse_gaussian}) we solve $\texttt{n-1}$ linear systems of size at most $\texttt{m}\times \texttt{m}$ where $m = \underset{i}{\max}\;|\texttt{N(i)}|$. This can be performed in $O(nm^3)$ flops. %, whereas the earlier pseudocode in (\ref{eq: pseudocode_full_gaussian}) for the dense model required $O(n^3)$ flops. 
Furthermore, these computations can be performed in parallel as each iteration of the loop is independent of the others. The above discussion provides a very useful strategy for constructing a sparse precision matrix. % from a dense precision matrix. 
Starting with a dense $n \times n$ matrix $\bC$, %and $\bC^{-1}$ both be dense $n\times n$ positive definite matrices. Suppose 
we construct a sparse strictly lower-triangular matrix $\bA$ with no more than $m (\ll n)$ non-zero entries in each row, and the diagonal matrix $\bD$ using the pseudocode in (\ref{eq: pseudocode_sparse_gaussian}) such that the matrix $\tildebC = (\bI-\bA)^{-1}\bD(\bI-\bA)^{-\top}$ is a covariance matrix whose inverse $\tildebC^{-1} = (\bI-\bA)^{\T}\bD^{-1}(\bI-\bA)$ is sparse. Figure~\ref{fig: NNGP_Chol} presents a visual representation of the sparsity. 

The factorization of $\tC^{-1}$ facilitates cheap computation of quadratic forms $\bu^\top \tC^{-1}\bv$ in terms $\bA$ and $\bD$. The algorithm to evaluate such quadratic forms \texttt{qf(u,v,A,D)} is provided in the following pseudocode:
\begin{align}\label{eq:nngp_qf}
\begin{array}{l}
\texttt{qf(u,v,A,D) = u[1]}\ast\texttt{v[1] / D[1,1]}\\
\texttt{for(i in 2:n)} \texttt{ \{ } \\ 
\qquad \texttt{qf(u,v,A,D) = qf(u,v,A,D) + (u[i] - dot(A[i,N(i)], u[N(i)]))} \\
\qquad \qquad \ast \texttt{(v[i] - dot(A[i,N(i)], v[N(i)]))/D[i,i]} \\
\qquad \texttt{\},}
\end{array}
\end{align}
where $*$ and $/$ denote multiplication and division by scalars, respectively. Observe (\ref{eq:nngp_qf}) only involves inner products of $m \times 1$ vectors. So, the entire \texttt{for} loop can be computed using $O(nm)$ flops as compared to $O(n^2)$ flops typically required to evaluate quadratic forms involving an $n \times n$ dense matrix. Also, importantly, the determinant of $\tC$ is obtained with almost no additional cost: it is simply $\prod_{\texttt{i=1}}^{\texttt{n}}\texttt{\bD[i,i]}$.

Hence, while $\tildebC$ need not be sparse, the density $N(\bw\given \bzero, \tildebC)$ is cheap to compute requiring only $O(n)$ flops. % since $\tildebC^{-1}$ is sparse and $\det(\tildebC)$ is the product of the diagonal elements of $\bD$. 
This was exploited by \cite{nngp} where the neighbor sets were constructed based on $m$ nearest neighbors and the traditional GP prior for $\bw$ in (\ref{eq:likewy}) was replaced with an NNGP prior $N(\bw \given \bzero, \tC)$. 
 %where $\tC$ is the NNGP precision matrix for the locations $\{\bs_1,\bs_2, \ldots, \bs_n\}$ i.e., the hierarchical NNGP model is now specified as 
%\begin{equation}\label{eq:nngplikewy}
%N(\by \given \bX\bbeta + \bw, \taus \bI) \; N(\bw \given \bzero, \tildebC) 
%\end{equation}
%The structure of the NNGP precision matrix $\tC^{-1}$ helps mitigate all of the aforementioned computational issues with the full GP model (\ref{eq:likewy}). Construction of $\tC^{-1}$ only requires storage of $n$ small $m \times m$ matrices. Also $\tC^{-1}$ is sparse with $O(nm^2)$ entries where $m$ denotes the number of nearest neighbors used. Furthermore, $\tC^{-1}$ and $\det(\tC)$ can be evaluated using 
%$O(nm^3)$ flops. Also, predicting at $K$ new locations require $O(Km^3)$ additional flops. These properties ensure that the MCMC algorithm for the hierarchical NNGP model (\ref{eq:nngplikewy}) enjoys computational and storage requirements that scale linearly with $n$ --- the size of the dataset. %NNGP, being a proper Gaussian Process enables it to be seamlessly embedded in any hierarchical spatial regression setting and allows for full inference of the latent spatial random effects $\bw$. This is often of scientific interest for researchers, as the residual spatial surface $w(\bs)$ may offer insight on local trends or the effect of unobserved covariates. 
The Markov chain Monte Carlo (MCMC) implementation of the NNGP model in \cite{nngp} requires updating the $n$ latent spatial effects $\bw$ sequentially, in addition to the regression and covariance parameters. %The Gibbs' sampler Algorithm used in \cite{nngp} updates each spatial effect $w(\bs)$ sequentially. 
While this ensures substantial computational scalability in terms of evaluating the likelihood, the behavior of MCMC convergence for such a high-dimensional model is difficult to study and may well prove unreliable. 

We observed that, for very large spatial datasets, sequential updating of the random effects often leads to very poor mixing in the MCMC (see Figures \ref{syn-large-gr} and \ref{syn-large-seq-chains}). The computational gains per MCMC iteration is thus offset by a slow converging MCMC. \citet{liu94} showed that MCMC algorithms where one or more variables are marginalized out tend to have lower autocorrelation and improved convergence behavior. Here we explore NNGP models that drastically reduce the parameter dimensionality of the NNGP models by marginalizing over the entire vector of spatial random effects. Three different variants are developed, including an MCMC free conjugate model, and their relative merits and demerits are assessed both in terms of computational burden as well as model prediction and inference. Simulation experiments using spatial datasets of up to 10 million locations are conducted to assess the models' performance. Finally, we use the NNGP models to analyze the TIU dataset comprising over 5 million locations. To our knowledge, fully Bayesian analysis of spatial data at such scales is unprecedented.
% cannot be achieved using traditional Gaussian Process based spatial regression models become computationally infeasible due to their

\subsection{Collapsed NNGP}\label{sec:nngpcol} %The NNGP model for the response is very convenient if the statistical objective is kriging at new locations and estimation of the regression coefficients $\bbeta$. However, as discussed earlier, 
\noindent The hierarchical model (\ref{eq:likewy}) or its NNGP analogue impart a nice interpretation to the spatial random effects. The latent surface $w(\bs)$ can provide a lot of information about the effect of missing covariates or unobserved physical processes. Hence, inference about $\bw$ is often critical for the researchers in order to improve the understanding of the underlying scientific phenomenon. 
%\cite{nngp} demonstrated that a direct NNGP model for the response will often render the post-hoc recovery of random effects impossible. 
Here, we provide a collapsed NNGP model that enjoys the frugality of a low-dimensional MCMC chain but allows for full recovery of the latent random effects.  %such that the $(i,j)^{th}$ element of $C[A,B]$ denotes the covariance $C(\ba_i,\bb_j \given \theta)$
We begin with the two-stage hierarchical specification $N(\by \given \bX\bbeta + \bw, \taus \bI) \times N(\bw \given \bzero, \tildebC) $ %and replace the GP prior for the random effects $w(\bs)$ with the NNGP prior $NNGP(0,\widetilde C(\cdot,\cdot))$. So, $\bw \sim N(\bzero, \tC)$ where $\tC$ is the $n \times n$ NNGP covariance matrix derived from $\bC$. 
and avoid sampling $\bw$ in the Gibbs' sampler by integrating out $\bw$ to obtain the collapsed NNGP model
\begin{equation}\label{eq:collapse}
\by \sim N(\bX\bbeta, \bLam) \mbox{ where } \bLam = \tC+\taus\bI
\end{equation}
This model has only $p+4$ parameters compared to $n+p+4$ parameters in the hierarchical model. We use a conjugate prior $N(\bmu_\beta,\bV_\beta)$ for $\bbeta$, Inverse Gamma priors for the spatial and noise variances, and uniform priors for the range and smoothness parameters. We use the $u\given \cdot$ notation to denote the full conditional distribution of any random variable $u$ in the Gibbs' sampler. Let $N(i)$ denote the set of indices corresponding to neighbor set of $\bs_i$. %and for any vector $\bv$, $\bv[N(i)]$ denote the subvector of $\bv$ corresponding  to the indices in $N(i)$. 
%Also, for any two sets of of locations $A=\{\ba_1,\ba_2, \ldots, \ba_k\}$ and $B=\{\bb_1,\bb_2,\ldots,\bb_l\}$, $\bC(A,B)$ denotes the $k \times l$ cross covariance matrix $((C(\ba_i,\bb_j \given \btheta))$ where $C(\cdot, \cdot \given \btheta)$ is the full GP covariance specified in (\ref{eq:matern}). For any subset of locations $S \subset \{\bs_1,\bs_2,\ldots,\bs_n\}$, $\by_S$ denotes the subvector of $\by$ corresponding to the indices representing $S$. Similar notation is used to denote $\bw_S$ and $\bX_S$. 
Observe that, although from Section~\ref{sec:nngp} we know $\tC=(\bI-\bA)^{-1}\bD(\bI-\bA)^{-\top}$, $\bLam$ does not enjoy any such convenient factorization. In fact, $\bLam^{-1}$ is also not guaranteed to be sparse, but exploiting the Sherman Woodbury Morrison (SWM) identity, we can write $\bLam^{-1}=\tau^{-2}\bI - \tau^{-4}\bOmega^{-1}$ where $\bOmega=(\tC^{-1}+\tau^{-2}\bI)$ enjoys the same sparsity as $\tC^{-1}$. Also, using a familiar determinant identity, we have $\det(\bLam) = \tau^{2n} \det(\tC) \det(\bOmega)$. 

We exploit these matrix identities in conjunction with sparse matrix algorithms to obtain posterior distributions of the parameters $\{\bbeta,\btheta,\taus\}$. In fact, the necessary computations can be done by entirely avoiding expensive matrix computations and is described in detail in Algorithm \ref{alg:col}. In addition to the inner product function $\texttt{dot}(\cdot,\cdot)$ introduced earlier, we require a fill-reducing permutation matrix and a sparse Cholesky factorization ($\texttt{sparsechol}(\cdot)$) for a sparse positive-definite matrix. Large matrix-matrix and matrix-vector multiplications either involve at least one triangular matrix ($\texttt{trmm}(\cdot,)$ or $\texttt{trmv}(\cdot,\cdot)$) or at least one sparse matrix ($\texttt{sparsemm}(\cdot,\cdot)$ or $\texttt{sparsemv}(\cdot,\cdot)$). We also use $\texttt{diagsolve}(\cdot,\cdot)$ and $\texttt{trsolve}(\cdot,\cdot)$ to solve linear systems with a diagonal or triangular coefficient matrix, respectively. We perform Cholesky decompositions, matrix-vector multiplications and solve linear equations involving general unstructured matrices using $\texttt{chol}(\cdot)$, $\texttt{gemv}(\cdot,\cdot)$ and $\texttt{solve}(\cdot,\cdot)$, respectively, only for small $p\times p$ or $m \times m$ matrices where both $p$ and $m$ are much less than $ n$. Other utilities used in Algorithm~\ref{alg:col} are $\texttt{diag}(\cdot)$ to extract the diagonal elements of a matrix, $\texttt{prod}(\cdot)$ to compute the product of the elements in a vector and $\texttt{rnorm}(\cdot)$ to generate a specified number of random variables (as an integer argument) from a standard $N(0,1)$ distribution.      
%The MCMC algorithm also recovers the latent random effects $w(\bs_i)$ for inference and provides prediction at new locations.

\begin{singlespacing}
{\small
\begin{breakablealgorithm}
\caption{Collapsed NNGP: Sampling from the posterior}\label{alg:col}
\begin{algorithmic}[1]
\makeatletter
%\Require something
%\Ensure something\vskip 3mm 
\Algphase{MCMC steps for updating $\{\bbeta,\btheta,\taus\}$} 
%\Statex Steps \hfill Flops
\State \textbf{Gibb's sampler update for $\bbeta$:}
\Statex $\bbeta \given \cdot \sim N(\bB^{-1}\bb, \bB^{-1})$, where $\bB= \bX^\top\bLam^{-1}\bX+\bV_\beta^{-1}$ and $\bb = \bX^\top\bLam^{-1}\by+\bV_\beta^{-1}\bmu_\beta$
\begin{enumerate}[(a)]
\item Use (\ref{eq: pseudocode_full_gaussian}) to obtain  $\texttt{\bA}$ and $\texttt{\bD}$ using $\texttt{\bC}$ and $\{N(i) \given i=1,2,\ldots,n\}$ \hfill $O(nm^3)$ flops
\item \texttt{$\Omega = \texttt{trmm}((\bI-\bA)^\top,\texttt{diagsolve}(\bD,\bI - \bA))+\tau^{-2} \ast \bI$} \hfill $O(nm^2)$ flops 
%\item $\Omega\texttt{=trmm((\bI-\bA)$^\top$, diagsolve(\bD,\bI - \bA))+ }\tau^{-2} \ast \texttt{ \bI}$ \hfill $O(nm^2)$ flops
\item Find a fill reducing permutation matrix $\texttt{\bP}$ for $\Omega$ %\hfill \textbf{flops?}
\item \texttt{$\bL = \texttt{sparsechol}(\texttt{sparsemm}(\texttt{sparsemm}(\bP,\Omega),\bP^{\top}))$} %\hfill \textbf{flops?}
%\item $ \texttt{\bL = sparchol(\bP\;}\%\ast\%\;\Omega\;\%\ast\%\;\texttt{\;\bP}^\top\texttt{)}$
\item $\texttt{ for (j in 1:n) \{  }$
\Statex \texttt{$\qquad \texttt{\bu}_j = \texttt{trsolve}(\bL,\texttt{sparsemv}(\bP,\texttt{\bX[,j]}))\;;\quad \texttt{\bv}_j = \texttt{trsolve}(\bL^{\top}, \bu_j)$}
\Statex $\quad \texttt{\}}$
% \item $\texttt{ for (j in 1:n) \{  }$
% \Statex $\qquad \texttt{\bu}_j\texttt{ = trsolve(\bL,\bP}\%\ast\%\texttt{\bX[,j])}$
% \Statex	$\qquad \texttt{\bv}_j\texttt{ = trsolve(\bL}^\top\texttt{,\bu}_j\texttt{)}$
% \Statex $\qquad \texttt{\}}$
\item \texttt{$\bF = \texttt{solve}(\bV_\beta,\bI)\;;\quad \texttt{f} = \texttt{solve}(\texttt{V}_\beta, \bmu_{\beta})$} \hfill $O(p^3)$ flops
%\item $\texttt{\bF = inv(\bV} _\beta\texttt{)} $  \hfill $O(p^3)$ flops
%\Statex $\texttt{f= solve(V}_\beta,\mu_\beta\texttt{)}$ 
\item Solve for $p\times p$ matrix $\bB$ and $p\times 1$ vector $\bb$: \hfill $O(np^2)$ flops
\Statex \texttt{$\texttt{for (j in 1:p)}$ \{ }
%\Statex $\texttt{ for (j in 1:p) \{ }$
\Statex \texttt{$\qquad \quad \texttt{\bb[j]} = \texttt{dot}(\texttt{y}, \texttt{X[,j]})/\taus - \texttt{dot}(\texttt{y}, \texttt{sparsemv}(\texttt{P}, \bv_j))/\tau^4 + \texttt{f[j]}$}
%\Statex $\qquad \; \; \texttt{\bb[j]=y}^\top\%\ast\%\texttt{X[,j]}/\taus -  \texttt{y}^\top\%\ast\%\texttt{\bP}\%\ast\%\texttt{\bv}_j/\tau^4 \texttt{+ f[j] }$
\Statex $\qquad \quad \texttt{for (i in 1:p) \{ }$
\Statex $\qquad \quad \qquad \texttt{B[i,j]} = \texttt{dot}(\texttt{X[,i]}, \texttt{X[,j]})/\taus - \texttt{dot}(\texttt{X[,i]}, \texttt{sparsemv}(\texttt{P}, \bv_j))/\tau^4 + \texttt{F[i,j]}$
%\Statex $\qquad \qquad \texttt{B[i,j]=X[,i]}^\top\%\ast\%\texttt{X[,j]}/\taus -  \texttt{X[,i]}^\top\%\ast\%\texttt{\bP}\%\ast\%\texttt{\bv}_j/\tau^4 \texttt{+ F[i,j] }$
\Statex $\qquad \qquad \texttt{\}}$
\Statex $\texttt{\}}$
%\item Solve the triangular systems $\bL\bu_j=\bP\bx_j$ and $\bL^\top\bv_j=\bu_j$ to obtain $\bP^\top\bv_j=(\tC^{-1}+\tau^{-2}\bI)^{-1}\bx_j$ for $j=1,2,\ldots,p$
%\item Using Sherman Woodbury Morrison (SWM) identity, calculate the quadratic forms $\bx_i^\top\bLam^{-1}\bx_j=\tau^{-2} \bx_i^\top \bx_j - \tau^{-4} \bx_i^\top \bP^\top\bv_j$ for all $i,j$ to obtain $\bX^\top\bLam^{-1}\bX$ \hfill $O(np^2)$ flops
%\item Calculate $\bX'\bLam^{-1}\by$ using SWM, similar to Steps 1(d) - (e) \hfill $O(np)$ flops
%\item Calculate $\bLam_\beta$ and $\bX'\bLam^{-1}\by+\bV_\beta^{-1}\bmu_\beta$ to generate from $\bbeta \given \cdot$ \hfill $O(p^3)$ flops
%\item Generate $\bbeta \given \cdot \sim \texttt{N(solve(B,b),inv(B))}$ \hfill $O(p^3)$flops
\item $\bbeta = \texttt{solve(B,b) + trsolve(chol(B),rnorm(p))}$ \hfill $O(p^3)$ flops
\end{enumerate}
%\Statex
\State \textbf{Metropolis-Hastings (MH) update for $\{\btheta,\taus\}$:}
\Statex $\displaystyle p(\btheta,\taus \given \cdot) \propto p(\btheta,\taus) \times \frac {1}{\sqrt{\det(\bLam)}} \exp\left(- \frac 12 (\by -\bX\bbeta)^\top\bLam^{-1}(\by-\bX\bbeta) \right)$
\begin{enumerate}[(a)]
	\item \texttt{$ \br = \by - \texttt{gemv}(\bX,\bbeta)\;;\; \bu = \texttt{trsolve}(\bL, \texttt{sparsemv}(\bP, \br))\;;\; \bv = \texttt{trsolve}(\bL^{\top},\bu)$} \hfill $O(np)$ flops
%	\item $\texttt{\bu = trsolve(\bL,\bP}\%\ast\%\texttt{(\by-\bX}\%\ast\%\beta\texttt{))}$
%	\Statex $\texttt{\bv= trsolve(\bL}^\top\texttt{,\bu)}$
	\item \texttt{$\texttt{q} = \texttt{dot}(\br,\br)/\taus - \texttt{dot}(\br, \texttt{sparsemv}(\bP,\bv))/\tau^4 $} \hfill %\textbf{flops?}
%	\Statex $\texttt{q = r}^\top\%\ast\%\texttt{r /}\taus \texttt{ - r}^\top\%\ast\%\texttt{ \bP }\%\ast\%\texttt{\bv}/\tau^4$ 
	\item \texttt{$\texttt{d} = \tau^{2*n}\ast \texttt{prod(diag(D))} \ast \texttt{prod(diag(L))}^2$} \hfill $O(n)$ flops
%	\item $\texttt{d=}\tau^{2*n} \ast \texttt{prod(diag(D))} \ast \texttt{prod(diag(L))}^2 $ \hfill $O(n)$ flops
%	\item    Initialize $\texttt{d=}\tau^{2*n}$ \hfill $O(n)$ flops
%	\Statex $\texttt{ for (i in 1:n) \{ }$ 
%	\Statex $\qquad \texttt{d=d}\ast \texttt{D[i,i]}\ast\texttt{L[i,i]}^2$ 
%	\Statex $\qquad \texttt{\}}$
	\item Generate $\displaystyle \texttt{p}(\btheta,\taus\given\cdot) \propto \frac{\texttt{exp}(\texttt{-q/2})\ast \texttt{p}(\btheta,\taus)}{\texttt{sqrt}(\texttt{d})}$
%	\item Generate $\texttt{p(}\theta \given \cdot\texttt{)} \propto \texttt{1/sqrt(d)} \ast \texttt{exp(-q/2)} \ast \texttt{p}(\theta)$ \hfill $O(1)$ flops
\end{enumerate}
%\Statex (a)  Solve the triangular systems $\bL\bu=\bP(\by-\bX\bbeta)$ and $\bL'\bv=\bu$ to obtain $\bP'\bv=(\tC^{-1}+\tau^{-2}\bI)^{-1}(\by-\bX\bbeta)$
%\Statex (b) Calculate $(\by -\bX\bbeta)'\bLam^{-1}(\by-\bX\bbeta)$ using SWM, similar to Step 1(d) \hfill $O(np)$ flops
%\Statex (c) Calculate $\det(\tildebC)$ and $\det(\tC^{-1}+\tau^{-2}\bI)=\det(\bL)^2=\prod_{i=1}^n L_{ii}^2$ \hfill $O(n)$ flops
%\Statex (d) Using matrix determinant identity, calculate $\det(\bLam) = \tau^{2n} \det(\tC) \det(\tC^{-1}+\tau^{-2}\bI)$
%\Statex (e) Using steps (b) and (d), generate from $p(\btheta \given \cdot) $ using MH
%\Statex 
\State Repeat Steps (1) and (2) (except Step 1(c)) $N$ times to obtain $N$ MCMC samples for $\{\bbeta,\btheta,\taus\}$
\end{algorithmic}
\end{breakablealgorithm}
}
\end{singlespacing}

%Here \texttt{prod(v)} gives the product of all the entries of a vector \texttt{v}, \texttt{diag(M)} denotes the vector consisting of the diagonal entries of a matrix \texttt{M}, \texttt{inv(M)}  and \texttt{chol(M)} respectively denote  the inverse and Cholesky factor of a matrix \texttt{M}, \texttt{sparchol(M)} is the Cholesky factor of a sparse matrix \texttt{M}, \texttt{trsolve(M,y)} and \texttt{diagsolve(M,y)} are the solutions to the linear system \texttt{Mx=y} where \texttt{M} is respectively a triangular and a diagonal matrix, \texttt{rnorm(n)} generates a $n \times 1$ vector of independent and identically distributed standard normal random variables and \texttt{Matern(s,t,}$\theta$\texttt{)} returns the value of the function in (\ref{eq:matern}) evaluated at location \texttt{s} and \texttt{t} and parameters $\theta$. 

Observe that the entire Algorithm~\ref{alg:col} is devoid of any expensive operations like \texttt{solve}, \texttt{chol} or \texttt{gemv} on dense $n \times n$ matrices. All such operations are limited to $m \times m$ or $p \times p$ matrices, where both $m$ and $p$ are small. The computational costs in terms of flops of all such steps are listed in the algorithm and are linear in $n$. However, the exact cost of the steps involving $\bL$ in Algorithm~\ref{alg:col} (Steps 1(c)-(e)) depends on the data design. Although $\bOmega$ is sparse $O(nm^2)$ non-zero entries, the sparsity of its Cholesky factor $\bL$ actually depends on the location of the non-zero entries. Hence we used a fill reducing permutation $\bP$ that increases the sparsity of the Cholesky factor. Although $\bP$ needs to be evaluated only once before the MCMC, finding the optimal $\bP$ yielding the least fill-in is an NP-complete problem. Hence algorithms have been proposed to improve sparsity patterns based on a variety of fill-in minimizing heuristics, see, e.g., \citet{amestoy96}, \citet{karypis98}, \citet{hager02} (also see Section~\ref{sec:sim}).

When flops per iteration of MCMC are considered, computational requirements for the collapsed NNGP model is data dependent and may exceed the exact linear flops usage for the hierarchical NNGP Algorithm. We also observed this in simulation experiments described in Section \ref{sec:sim}. However, the improved MCMC convergence for the collapsed NNGP, as observed in Figures \ref{syn-large-gr} and \ref{syn-large-col-chains}, implies that substantial computational gains accrue by truncating the MCMC run. Furthermore, all the \texttt{for} loops in Algorithm \ref{alg:col} can be evaluated independent of each other using parallel computing resources.

The collapsed model nicely separates the MCMC sampler for parameter estimation from posterior estimation of spatial random effects and subsequent predictions. %We observed in Figure \ref{syn-large-gr} that the reduced parameter dimensionality significantly improved the MCMC convergence.
Computational benefits accrue from using the quantities $\bL$ and $\bu$ already computed in Steps~1(d)~and~2(a) of Algorithm~\ref{alg:col} corresponding to the post-convergence samples of $\{\bbeta,\btheta,\tau^2\}$. This is presented in the algorithm below.  

\begin{singlespacing}
{\small
\begin{breakablealgorithm}
\caption{Collapsed NNGP: Posterior predictive inference}\label{alg:col_post_mcmc}
\begin{algorithmic}[1] 
\Algphase{Post-MCMC steps using $\bL$ and $\bu$ from Steps~1(d)~and~2(a) of Algorithm~\ref{alg:col} for post-convergence samples of $\{\bbeta,\btheta,\taus\}$}
\State \textbf{Sample from $p(\bw \given \cdot)$ one-for-one for each post-convergence sample of $\{\bbeta,\btheta,\taus\}$}
\Statex $\bw \given \cdot \sim N(\bB^{-1}\bb, \bB^{-1})$, where $\bB = \tC^{-1} + \tau^{-2} \bI$ and $\bb=(\by-\bX\bbeta)/\taus$
\begin{enumerate}[(a)]
	\item $\texttt{z} = \texttt{rnorm}(n)$ \hfill $O(n)$ flops
%	\item  $\texttt{z= rnorm(n)}$ \hfill $O(n)$ flops
	\item %\texttt{$\br = \by - \texttt{gemv}(\bX,\bbeta)\;;\quad \texttt{\bu} = \texttt{trsolve}(\bL,\texttt{sparsemv}(\bP,\texttt{r}))$} \hfill \textbf{flops?}  
%	$\texttt{\bu = trsolve(\bL,\bP}\%\ast\%\texttt{(\by-\bX}\%\ast\%\beta\texttt{))}$
	%\Statex 
	\texttt{$\bw = \texttt{sparsemv}(\bP^{\T},\texttt{trsolve}(\bL^{\top},\bu/\taus + \bz))$} %\hfill \textbf{flops?}
%	\Statex $\texttt{w= \bP}^\top \%\ast\% \texttt{trsolve(\bL}^{\top}\texttt{,\bu}/\taus \texttt{+ \bz )}$ 
\end{enumerate}
%\Statex
\State \textbf{Prediction at a new location $\bs_0$:}
\Statex $y(\bs_0) \given \cdot \sim N(\bx(\bs_0)^\top\bbeta + w(\bs_0), \taus)$
\begin{enumerate}[(a)]
\item Find $\texttt{N}_0$ --- set of $m$ nearest neighbors of $\bs_0$ among $\{\bs_1,\bs_2,\ldots,\bs_n\}$ \hfill $O(n)$ flops
\item \texttt{$\texttt{c} = \texttt{C}(\bs_0, \texttt{N}_0;\btheta)$} \hfill $O(m)$ flops
%\item $\texttt{c= Matern(s}_0\texttt{,N(0)},\theta\texttt{)}$ \hfill $O(m)$ flops
\item \texttt{$\texttt{m} = \texttt{dot}(\texttt{c}, \texttt{solve}(\texttt{C}(\texttt{N}_0,\texttt{N}_0), \texttt{w}[\texttt{N}_0]))$} \hfill $O(m^3)$ flops
%\item $\texttt{m= c }^\top\%\ast\% \texttt{solve(C[N(0),N(0)],w[N(0)])}$ \hfill $O(m^3)$ flops
\Statex \texttt{$\texttt{v} = \texttt{C}(\bs_0,\bs_0;\btheta) - \texttt{dot}(\texttt{c}, \texttt{solve}(\texttt{C}(\texttt{N}_0,\texttt{N}_0), \texttt{c}))$}
%\Statex $\texttt{v= } \sigs \texttt{- c }^\top\%\ast\% \texttt{solve(C[N(0),N(0)],c)}$
\item \texttt{$\texttt{w}(\bs_0) = \texttt{m} + \texttt{sqrt}(\texttt{v})\ast \texttt{rnorm}(1)$} \hfill $O(p)$ flops 
%\item $\texttt{w0 = m + sqrt(v)} \ast \texttt{rnorm(1)}$ \hfill $O(p)$ flops
\Statex \texttt{$\texttt{y}(\bs_0) = \texttt{dot}(\texttt{x}(\bs_0),\bbeta) + \texttt{w}(\bs_0) + \tau\ast\texttt{rnorm}(1)$} \hfill $O(p)$ flops
%\Statex $\texttt{y0 = x0}^\top \%\ast\% \beta \texttt{ + w0 + } \tau  \ast \texttt{ rnorm(1)}$ 
%\Statex (b) Calculate $\bc_0= \bC(\{\bs_0\}, N(\bs_0))$ and $\bC_0= \bC(N(\bs_0), N(\bs_0))$ \hfill $O(m^2)$ flops
%\Statex (c) For each post burn-in sample of $\bw$ and $\btheta$, generate $w(\bs_0) \given \cdot \sim N(\bc_0'\bC_0^{-1}\bw_{N(\bs_0)}, \sigs - \bc_0'\bC_0^{-1}\bc_0)$ \hfill $O(m^3)$ flops
%\Statex (d) For each post burn-in sample of $\bw(\bs_0)$, $\bbeta$ and $\btheta$, generate $\by(\bs_0) \given \cdot \sim N(\bx(\bs_0)'\bbeta+w(\bs_0), \taus)$ \hfill $O(p)$ flops
\end{enumerate}
 \end{algorithmic}
\end{breakablealgorithm}
}
\end{singlespacing}
\noindent Algorithm~\ref{alg:col_post_mcmc} demonstrates how inference on $w(\bs)$ and $y(\bs)$ can be easily achieved for any spatial location using the post burn-in samples of $\{\bbeta,\btheta,\taus\}$. We first sample the spatial random effects $p(\bw\given\by)$ for the observed locations, use them to sample from $p(w(\bs_0)\given \by)$ and then from $p(y(\bs_0)\given \by)$.

%Algorithm 1 also provides the computational costs in terms of flops for some of the steps involved in terms of the sample size $n$, number of predictors $p$ and number of nearest neighbors $m$. However, the flops for some of the intermediate steps depend on the actual spatial locations. This is because, even though $\tC^{-1}$ is sparse with $O(nm^2)$ entries, the sparsity of the Cholesky factor of $\tC^{-1}+\tau^{-2}\bI$ actually depends on the sparsity pattern of $\tC^{-1}$, which is governed by the spatial locations. %If a fill-in reducing step is not used, $\bP$ in all subsequent steps should be replaced by the identity matrix.

\subsection{NNGP for the response}\label{sec:nngpy} 
\noindent Both the sequential NNGP Algorithm in \cite{nngp} or the collapsed version in Section \ref{sec:nngpcol} accomplishes prediction at a new location via recovering the spatial random effects first, proceeded by kriging at the new location. This differed from \cite{ve88}'s original approach which applied nearest neighbor approximation directly to the marginal likelihood of $\by$. The recovery of the spatial random effects becomes necessary if inference on the latent process is of interest.  Although recovering $\bw$, as discussed earlier, has its own importance, if spatial interpolation of the response is the primary objective, this intermediate step is often a computational burden. In this Section, we propose a NNGP model for the response $\by$ that sacrifices the ability to recover $\bw$ and directly predicts the response at new locations. 

\cite{nngp} demonstrated that an NNGP model can be derived from any Gaussian Process. If $w(\bs) \sim GP(0,C(\cdot,\cdot))$ then the response $y(\bs) \sim GP(\bx(\bs)^\top\bbeta, \Sigma(\cdot,\cdot))$ is also a Gaussian Process where $\Sigma(\bs_i,\bs_j)=C(\bs_i,\bs_j)+\taus I(\bs_i=\bs_j)$. % is the covariance function created by adding a white noise  to $C(\cdot,\cdot)$. 
Hence, we can directly derive an NNGP for the response  process $y(\bs)$. For finite dimensional realizations $\by$, likelihood under the response NNGP model is identical to Vecchia's composite likelihood. \cite{nngp} extend this notion to a fully Bayesian setup. The key observation is that Vecchia's approximation corresponds to a proper multivariate Gaussian distribution obtained by simply replacing the covariance matrix $\bSig = \bC + \taus \bI$ with its nearest-neighbor approximation $\tSig$ as described in Section~\ref{sec:nngp}. The sparsity properties documented in Section~\ref{sec:nngp} apply to $\tSig$ as well. MCMC steps for parameter estimation and prediction using this response NNGP model are provided in Algorithm~\ref{alg:resp}.

\begin{singlespacing}
{\small 
\begin{breakablealgorithm}
	\caption{Response NNGP model: Sampling from the posterior}\label{alg:resp}
	\begin{algorithmic}[1]
		%\Require something
		%\Ensure something
		\vskip 3mm \Algphase{MCMC steps for updating $\{\bbeta,\btheta,\taus\}$} 
		%\Statex Steps \hfill Flops
		\State \textbf{Gibb's sampler update for $\bbeta$:}
		\Statex $\bbeta \given \cdot \sim N(\bB^{-1}\bb,\bB^{-1})$, where $\bB =\bX^\top\tSig^{-1}\bX+\bV_\beta^{-1}$ and $\bb = \bX^\top\tSig^{-1}\by+\bV_\beta^{-1}\bmu_\beta$
		\begin{enumerate}[(a)]
		\item Use (\ref{eq: pseudocode_full_gaussian}) to obtain  $\texttt{\bA}$ and $\texttt{\bD}$ using $\Sigma$ and $\{N(i) \given i=1,2,\ldots,n\}$ \hfill $O(nm^3)$ flops
		\item \texttt{$\bF = \texttt{solve}(\bV _\beta, \bI)\;;\quad \texttt{f} = \texttt{solve}(\bV_\beta,\bmu_\beta)$} \hfill $O(p^3)$ flops
%		\item $\texttt{\bF = inv(\bV} _\beta\texttt{)} $ \hfill $O(p^3)$ flops
%		\Statex $\texttt{f= solve(V}_\beta,\mu_\beta\texttt{)}$ 
		\item Solve for $p \times p$ matrix $\bB$ and $p \times 1$ vector $\bb$ using (\ref{eq:nngp_qf}):  \hfill $O(nmp^2)$ flops
		\Statex $\texttt{for (i in 1:p) \{}$
		\Statex $\qquad\;\; \texttt{b[i]} = \texttt{qf}(\texttt{\bX[,i],\by,\bA,\bD}) + \texttt{f}[i]$
%		\Statex $\qquad \; \; \texttt{b[i] = qf(X[,i],y,A,D) + f[i]}$
		\Statex $\qquad\;\; \texttt{for (j in 1:p) \{  }$	
		\Statex $\qquad \qquad \; \; \texttt{B[1,j]} = \texttt{qf}(\texttt{\bX\texttt{[,i]},\bX\texttt{[,j]},\bA,\bD)} + \texttt{\bF}[1,j]$
		\Statex $\qquad \qquad \texttt{\}}$
		\Statex $\qquad \; \ \texttt{\}}$
		\item $\bbeta = \texttt{solve(B,b)} + \texttt{trsolve}(\texttt{chol(B)},\texttt{rnorm(p)})$ \hfill $O(p^3)$ flops
		\end{enumerate}
		%\Statex
		\State \textbf{Metropolis-Hastings (MH) update for $\{\btheta,\taus\}$:}
		\Statex $p(\btheta,\taus \given \cdot) \propto  p(\btheta,\taus) \times \frac {1}{\sqrt{\det(\tSig)}} \exp\left(- \frac 12 (\by -\bX\bbeta)^\top\tSig^{-1}(\by-\bX\bbeta) \right)$
		\begin{enumerate}[(a)] 
		\item \texttt{$\texttt{e} = \by - \texttt{gemv}(\bX,\bbeta)\;;\quad \mbox{Using (\ref{eq:nngp_qf}), }\; \texttt{q} = \texttt{qf(e,e,\bA,\bD)}$} \hfill $O(n(p+m))$ flops
		%\Statex Using (\ref{eq:nngp_qf}), $\texttt{q} = \texttt{qf(e,e,\bA,\bD)}$
		%\item  Initialize $d= 1$
		%\Statex $\texttt{for(i in 1:n) d=d*D[i,i]}$ 
		\item \texttt{$\texttt{d} = \texttt{prod(diag(D))}$} \hfill $O(n)$ flops
		\item Generate \texttt{$\displaystyle \texttt{p}(\btheta,\taus\given \cdot) \propto \frac{\texttt{exp(-q/2)}\ast \texttt{p}(\btheta,\taus)}{\texttt{sqrt(d)}}$} \hfill $O(1)$ flops
		\end{enumerate}
		%\Statex 
		\State Repeat Steps (1) and (2) $N$ times to obtain $N$ MCMC samples for $\{\bbeta,\btheta,\taus\}$
		%\Statex
% 		\State \textbf{Prediction at a new location $\bs_0$ for each post burn-in sample of $\bbeta$ and $\btheta$:}
% 		\Statex $y(\bs_0) \given \cdot \sim N(\bx(\bs_0)^\top\bbeta+ \bc_0^\top\bSig_0^{-1}(\by_{N(\bs_0)}-\bX_{N(\bs_0)}\bbeta), \sigs+\taus - \bc_0^\top\bSig_0^{-1}\bc_0)$
% 		\begin{enumerate}[(a)]
% 		\item Find $N(\bs_0)$ --- set of $m$ nearest neighbors of $\bs_0$ among $\{\bs_1,\bs_2,\ldots,\bs_n\}$ \hfill $O(n)$ flops
% 		\item $\texttt{c= Matern(s}_0\texttt{,N(0)},\theta\texttt{)}$ \hfill $O(m)$ flops
% \item $\texttt{m= c }^\top\%\ast\% \texttt{solve(}\Sigma\texttt{[N(0),N(0)],y[N(0)]-X[N(0),]}\%\ast\%\beta\texttt{)}$ \hfill $O(m^3)$ flops
% \Statex $\texttt{v= } \sigs + \taus \texttt{- c }^\top\%\ast\% \texttt{solve(}\Sigma\texttt{[N(0),N(0)],c)}$
% \item $\texttt{y0 = x0}^\top \%\ast\% \beta \texttt{ + m +  sqrt(v) } \ast \texttt{ rnorm(1)}$ 
% 		\end{enumerate}
	\end{algorithmic}
\end{breakablealgorithm}
}
\end{singlespacing}
\noindent Unlike the collapsed NNGP model, the computational cost for each step of Algorithm~\ref{alg:resp} does not depend on the spatial design of the data and is exactly linear in $n$. This is a result of the complete absence of the latent spatial effects $\bw$ in the model. Once again, parallel computing can be leveraged to evaluate all the \texttt{for} loops. A caveat with the response model is that recovery of $\bw$ is not possible as highlighted in \cite{nngp}. However, if that is of peripheral concern, the response model offers a computationally parsimonious solution for fully Bayesian analysis of massive spatial datasets. Posterior predictive inference, therefore, consists only of predicting the outcome $y(\bs)$ at any arbitrary location $\bs$. This is achieved easily through Algorithm~\ref{alg:resp_post_mcmc} given below, where $\by_{N(\bs_0)}$ represents the subvector of $\by$ corresponding to the points in $N(\bs_0)$, $\bX_{N(\bs_0)}$ is the corresponding design matrix, and $\bSig_0$ is the $m\times m$ covariance matrix for $\by_{N(\bs_0)}$.

\begin{singlespacing}
{\small
\begin{breakablealgorithm}
\caption{Response NNGP model: Posterior predictive inference}\label{alg:resp_post_mcmc}
\begin{algorithmic}[1] 
\Algphase{Post-MCMC steps using post-convergence samples of $\{\bbeta,\btheta,\taus\}$}
\State \textbf{Sample from $p(y(\bs_0) \given \cdot)$ one-for-one for each post-convergence sample of $\{\bbeta,\btheta,\taus\}$}
\Statex $y(\bs_0) \given \cdot \sim N(\bx(\bs_0)^\top\bbeta + \bc_0^\top\bSig_0^{-1}(\by_{N(\bs_0)}-\bX_{N(\bs_0)}\bbeta), \Sigma(\bs_0,\bs_0) - \bc_0^\top\bSig_0^{-1}\bc_0)$
	\begin{enumerate}[(a)]
	\item Find $\texttt{N}_0$ --- set of $m$ nearest neighbors of $\bs_0$ among $\{\bs_1,\bs_2,\ldots,\bs_n\}$ \hfill $O(n)$ flops
	\item \texttt{$\texttt{c} = \Sigma(\bs_0,\texttt{N}_0;\btheta)$} \hfill $O(m)$ flops
	\item \texttt{$\texttt{m} = \texttt{dot}(\texttt{c},\texttt{solve}(\Sigma[\texttt{N}_0,\texttt{N}_0], \by[\texttt{N}_0]-\texttt{dot}(\bX[\texttt{N}_0,],\bbeta))$} \hfill $O(m^3)$ flops
	\Statex \texttt{$\texttt{v} = \Sigma(\bs_0,\bs_0) - \texttt{dot}(\texttt{c},\texttt{solve}(\Sigma[\texttt{N}_0,\texttt{N}_0],\texttt{c}))$}
	\item \texttt{$\texttt{y}(\bs_0) = \texttt{dot}(\bx(\bs_0),\bbeta) + \texttt{m} +  \texttt{sqrt(v)} \ast \texttt{rnorm(1)}$} \hfill $O(p)$ flops 
	\end{enumerate}
 \end{algorithmic}
\end{breakablealgorithm}
}
\end{singlespacing}

\subsection{MCMC-free exact Bayesian inference using conjugate NNGP}\label{sec:nngpex}
\noindent The fully Bayesian approaches developed in \cite{nngp} and in Sections \ref{sec:nngpcol} and \ref{sec:nngpy} provide complete posterior distributions for all parameters. However, for massive spatial datasets containing millions of observations, running the Gibbs' samplers for several thousand iterations may still be prohibitively slow. One advantage of NNGP over similar scalable statistical approaches for large spatial data is that it offers a probability model. Here, we exploit this fact to achieve exact Bayesian inference. 

We define $\alpha=\taus/\sigs$ and rewrite the marginal model from Section \ref{sec:nngpy} as $N(\by\given \bX\bbeta, \sigs \bM)$, where $\bM = \bG + \alpha \bI$ and $\bG$ denotes the Matern correlation matrix corresponding to the covariance matrix $\bC$ i.e. $\bG[i,j]=C(\bs_i,\bs_j,(1,\nu,\phi)^\top)$. Once again, the analogous NNGP model can be obtained by replacing the dense matrix $\bM$ with its nearest-neighbor approximation $\tM$. Note that $\tM$ depends on $\alpha$, the spatial range $\phi$ and smoothness $\nu$. Empirically, in spatial regression models, the spatial process parameters $\phi$ and $\nu$ are often not well estimated due to multimodality issues. In fixed domain asymptotic settings \citep[see, e.g.,][]{zhang04} it is impossible to jointly identify the spatial covariance parameters. Consequently, if inference for the covariance parameters is not of interest, it might be possible to fix them at reasonable values with minimal effect on prediction or point estimates of other model parameters. For example, the smoothness parameter $\nu$ could be fixed at $0.5$, which reduces (\ref{eq:matern}) to the exponential covariance function, and $\phi$ and $\alpha$ could be estimated using $K$-fold cross-validation. 

For fixed $\alpha$ and $\phi$, %the model for $\by$ is a standard linear regression where the covariance is known upto an unknown constant. % and evaluate $\bb_i$ and $f_i$ as in (\ref{Eq: bf}) with the fixed values $\sigs=1$, $\taus=\alpha$, $\phi$ and $\nu$. Let $\tSig^*$ denote the corresponding NNGP covariance matrix constructed using these $\bb_i$'s and $f_i$'s.
%Since $\tM$ is now a fixed known covariance matrix, the only unknown parameters now are $(\bbeta',\sigs)'$.
we obtain the familiar conjugate Bayesian linear regression model $\displaystyle IG(\sigs \given a_\sigma, b_\sigma) \times N(\bbeta \given \bmu_\beta, \sigs \bV_\beta) \times N(\by \given \bX\bbeta , \sigs \tM)$ with joint posterior distribution
\begin{align*}
p(\bbeta,\sigma^2\given \by) \propto \underbrace{IG(\sigma^2\given a_{\sigma}^*, b_{\sigma}^*)}_{p(\sigma^2\given \by)} \times \underbrace{N(\bbeta \given \bB^{-1}\bb, \sigma^2\bB^{-1})}_{p\left(\bbeta\given\sigma^2,\by\right)}\;,
\end{align*}
where $a_{\sigma}^* = a_{\sigma} + n/2$, $\displaystyle b_{\sigma}^* = b_\sigma+ \frac{1}{2}\left( \bmu_\beta^\top\bV^{-1}_\beta\bmu_\beta+\by^\top\tM^{-1}\by - \bb^\top \bB^{-1} \bb \right)$, $\bB = \bV^{-1}_\beta+\bX^\top\tM^{-1}\bX$ and $\bb = \bV_\beta^{-1}\bmu_\beta+\bX^\top\tM^{-1}\by$. %\textbf{[AD: Please verify and fill in the correct expressions for these.]} 
It is easy to directly sample $\sigma^2 \sim IG(a_{\sigma}^*, b_{\sigma}^*)$ and then sample $\bbeta \sim N(\bB^{-1}\bb,\sigma^2\bB^{-1})$ one-for-one for each drawn $\sigma^2$. This produces samples from the marginal posterior distributions $\displaystyle \bbeta \given \by \sim \mbox{MVS-}t_{2a^*_\sigma} \left(\bB^{-1}\bb, \frac {b^*_\sigma}{a^*_\sigma} \bB^{-1}\right)$ and $\displaystyle \sigs \given \by \sim IG(a^*_{\sigma},b^*_{\sigma})$,
% \begin{align}\label{eq:margpost}
% 	\bbeta \given \by & \sim \mbox{MVS-}t_{2a} \left(\bB^{-1}\bb, \frac {b}{a} \bB^{-1}\right)\;, \quad \sigs \given \by \sim IG(a^*_{\sigma},b^*_{\sigma})\;,
% \end{align} 
where $\mbox{MVS-}t_\kappa(\bB^{-1}\bb,(b/a)\bB^{-1})$ denotes the multivariate non-central Student's $t$ distribution with degrees of freedom $\kappa$, mean $\bB^{-1}\bb$ and variance $b\bB^{-1}/(a-1)$. The marginal posterior mean and variance for $\sigs$ are $b^*_\sigma/(a^*_\sigma-1)$ and $b^{*2}_\sigma/(a^*_\sigma-1)^2(a^*_\sigma-2)$, respectively. 
% \begin{eqnarray}\label{eq:nig}
% \bV=&(\bV^{-1}_\beta+\bX^\top\tM^{-1}\bX)^{-1} \nonumber \\
% \bg=&\bV (\bV_\beta^{-1}\bmu_\beta+\bX^\top\tM^{-1}\by) \\
% a=&a_\sigma + n/2 \nonumber  \\
% b=&b_\sigma+ \left( \bmu_\beta^\top\bV^{-1}_\beta\bmu_\beta+\by^\top\tM^{-1}\by - \bg^\top \bV^{-1} \bg \right)/2 \nonumber 
% \end{eqnarray}

Instead of sampling from the posterior directly, we prefer a fast evaluation of the marginal posterior distributions to effectively implement the aforementioned cross-validatory approach. A pseudocode for fast evaluation of the above are provided in Algorithm~\ref{alg:exact}. The marginal posterior predictive distribution at a new location $\bs_0$ is given by $y(\bs_0) \given \by \sim t_{2a^*_\sigma} (m_0, b^*_\sigma v_0 / a^*_\sigma)$ where expressions for $m_0$ and $v_0$ are provided in Step 3 of Algorithm \ref{alg:exact}.
%$m_0 = \bM[\bs_0,N(\bs_0)]\bM[N(\bs_0),N(\bs_0)]^{-1}\bB^{-1}\bb$ and $v_0 = ??$ \textbf{[AD: Please provide expressions for these here in terms of the quantities defined above.]}. 
We deploy hyper-parameter tuning based on $K$-fold cross-validation to choose the optimal $\alpha$ and $\phi$ from a grid of possible values. We denote the indices and locations corresponding to the $k$-{th} fold of the data by $I(k)$ and $S(k)$ respectively whereas $I(-k)$ and $S(-k)$ respectively denote the analogous quantities when the $k^{th}$ fold is excluded from the data. Also, let $N(i,k)$ denote the neighbor set for a location $\bs_i$ constructed from the locations in $S(-k)$. Details of the cross-validation procedure are also provided in Algorithm~\ref{alg:exact}. 

\begin{singlespacing}
{\small
\begin{breakablealgorithm}
	\caption{MCMC free posterior sampling for conjugate NNGP model}\label{alg:exact}
	\begin{algorithmic}[1]
		%\Require something
		%\Ensure something
		\vskip 3mm \Algphase{Hyper parameter tuning} 
		%\Statex Steps \hfill Flops
		%\State \textbf{Gibb's sampler update for $\bbeta$:}
		%\Statex $\bbeta \given \cdot \sim N(\bSig_\beta (\bX'\tSig^{-1}\by+\bV_\beta^{-1}\bmu_\beta), \bSig_\beta) \mbox{ where } \Sig_\beta=(\bX'\tSig^{-1}\bX+\bV_\beta^{-1})^{-1}$
		\State Fix $\alpha$ and $\phi$, split the data into $K$ folds. 
		\begin{enumerate}[(a)]
			  \item Find the collection of neighbor sets $ {\cal N} = \{ N(i,k) : i = 1,2,\ldots,n; k=1,2,\ldots,K\}$
%			\item \texttt{w= perm(1:n)}
%			\item \texttt{for (k in 1:K) \{}
%			\item 
		\end{enumerate}
		%\State Calculate 
		%\Statex
		\State Obtain posterior means for $\bbeta$ and $\sigs$ after removing the $k^{th}$ fold of the data:
		\begin{enumerate}[(a)]
			\item	Use (\ref{eq: pseudocode_full_gaussian}) to obtain $\texttt{\bA(k)}$ and $\texttt{\bD(k)}$ from $\bM[S(-k),S(-k)]$ and ${\cal N}$ \hfill $O(nm^3)$ flops
			\item \texttt{$\bF = \texttt{solve}(\bV_\beta, \bI)\;;\quad \texttt{\bff} = \texttt{solve}(\bV_\beta,\bmu_\beta) $} \hfill $O(p^3)$ flops
%			\Statex \texttt{$\texttt{f} = \texttt{solve}(\bV_\beta,\bmu_\beta)$} 
			\item Solve for $p \times p$ matrix $\bB(k)$ and $p \times 1$ vector $\bb(k)$ using (\ref{eq:nngp_qf}):  \hfill $O(nmp^2)$ flops
			\Statex $\texttt{for (i in 1:p) \{}$
			\Statex $\qquad \; \; \texttt{\bb(k)[i]} = \texttt{qf(\bX[S(-k),i],\by[S(-k)],\bA(k),\bD(k))} + \texttt{\bff[i]}$
			\Statex $\qquad \texttt{ for (j in 1:p) \{  }$	
			%		\Statex $\qquad \qquad \; \; \texttt{v[i]= v[i] + u[k] }\ast\texttt{ U[i,k] / D[i,i]}$
			\Statex $\qquad \qquad \; \; \texttt{\bB(k)[i,j]} = \texttt{qf(\bX[S(-k),i],\bX[S(-k),j],\bA(k),\bD(k))} + \texttt{\bF[i,j]}$
			\Statex $\qquad \texttt{\}}$
			\Statex $\texttt{\}}$
			\item  \texttt{$\texttt{\bV(k)} = \texttt{solve}(\bB(k),\bI)\;;\; \bg(k) = \texttt{gemv(\bV(k),\bb(k))}$} \hfill$O(p^3)$ flops
			%\Statex $\texttt{V(k)=inv(B(k))}$
			%\Statex $\texttt{g(k)= solve(B(k),v(k))}$
			\Statex \texttt{$\texttt{a$_{\sigma}^*$(k)} = \texttt{a}_{\sigma} + \texttt{(n-n/K)/2}$}
%			\Statex $\texttt{a(k)= a}_\sigma\texttt{ + (n-n/K)/2}$
			\Statex \texttt{$\texttt{b$_{\sigma}^*$(k)} = \texttt{b}_{\sigma} + (\texttt{dot}(\bmu_{\beta}, \bff) + \texttt{qf(\by[S(-k)],\by[S(-k)],\bA(k),\bD(k))} - \texttt{dot(\bg(k),\bb(k)))/2}$}
%			\Statex $\texttt{b(k)= b}_\sigma\texttt{ + (}\mu_\beta^\top \%\ast\% \texttt{f + qf(y[S(-k)],y[S(-k)],A(k),D(k)) - g(k)}^\top \%\ast\% \texttt{v(k))/2}$
			\item $\widehat \bbeta = \texttt{\bg(k)}\;;\;\quad \widehat\sigs = \texttt{b$_{\sigma}^*$(k)/(a$_{\sigma}^*$(k)-1)}$
%			\item $\widehat \bbeta= \texttt{g(k)}$; $\widehat \sigs = \texttt{b(k)/(a(k)-1)}$
		\end{enumerate}		
		%\Statex
		\State Predicting posterior means of $\by[S(k))]$: \hfill $O(nm^3/K)$ flops	
			\begin{enumerate}[(a)]
			\Statex $\texttt{for (s in S(k)) \{}$
			\Statex $\qquad \texttt{N(s,k)} = \mbox{m-nearest neighbors of $\bs$ from $\bS(-k)$}$
			\Statex $\qquad \texttt{\bz} = \texttt{M(\bs,N(\bs,k))}$ %\textbf{AD: Changed Matern to generic}
			\Statex $\qquad \texttt{\bw} = \texttt{solve(M[N(\bs,k),N(\bs,k)],z)}$
			\Statex $\qquad \texttt{m}_0 = \widehat{\texttt{y(\bs)}} = \texttt{dot(\bx(\bs), \bg(k))} + \texttt{dot(\bw,(\by[N(\bs,k)] - dot(\bX[N(\bs,k),],\bg(k))))}$
			\Statex \texttt{$\qquad \bu = \bx(\bs) - \texttt{dot(\bX[N(s,k),],\bw)}$}
			%\Statex $\qquad \texttt{u = x(\bs) - X[N(s,k),]}^\top \%\ast\% \texttt{w}$
			\Statex $\qquad \texttt{v}_0 = \texttt{dot(\bu, gemv(\bV(k),\bu))} + 1 + \alpha - \texttt{dot(\bw,\bz)}$
			%\Statex $\qquad \texttt{v}_y\texttt{ = u}^\top \%\ast\% \texttt{solve(B(k),u) + 1 + }\alpha\texttt{ - w}^\top \%\ast\%\texttt{z}$
			\Statex \texttt{$\qquad \widehat{\texttt{Var(y(\bs))}} = \texttt{b}_{\sigma}^{*}\texttt{(k)}\texttt{v}_0/(\texttt{a}_{\sigma}^{*}\texttt{(k)} - 1)$}
%			\Statex $\qquad \widehat{\texttt{Var(y(\bs))}}\texttt{ = 2}\ast \texttt{b(k) }\ast \texttt{v}_y\texttt{ / (2}\ast \texttt{a(k) - 1)} $
			\Statex $\texttt{\}}$
			%\Statex $\qquad \texttt{h = X[s,]-Matern(s,N(s,k),(1,}\phi\texttt{,0.5)}^\top\texttt{) \%\ast\% solve(M[N(s,k),N(s,k)],X[N(s,k),],z)}$
			%\item Find the RMSPE $\texttt{sum(y(S(k)-\widehat{\texttt{y(\bs)}}))}$
		\end{enumerate}
		%\Statex
		\State Root Mean Square Predictive Error (RMSPE) over $K$ folds: \hfill $O(n)$ flops
		\begin{enumerate}[(a)]
		\item Initialize $e=0$
		\Statex $\texttt{for (k in 1:K) for (s}_i\texttt{ in S[k]) \{} $
		\Statex $\qquad \texttt{e= e +(y(\bs}_i\texttt{)-}\widehat{\texttt{y(\bs}_i\texttt{)}}\texttt{ )}^2$
		\Statex $\texttt{\}}$
		\end{enumerate}
		%\Statex
		\State Cross validation for choosing $\alpha$ and $\phi$
		\Statex (a) Repeat steps (2) and (3) for $G$ values of $\alpha$ and $\phi$  \hfill $O(GKnm(p^2+m^2))$ flops
		\Statex (b) Choose $\alpha_0$ and $\phi_0$ as the value that minimizes the average RMSPE \hfill $O(G)$ flops
		\Algphase{Parameter estimation and prediction}
		\State Repeat step (2) with $(\alpha_0, \phi_0)^\top$ and the full data to get $(\bbeta,\sigs) \given \by$ \hfill $O(nmp^2+nm^3)$ flops
		%\Statex
		\State Repeat step (3) with $(\alpha_0, \phi_0)^\top$ and the full data to predict at a new location $\bs_0$ to obtain the mean and variance of  $y(\bs_0) \given \by$   \hfill $O(m^3)$ flops
		\end{algorithmic}
\end{breakablealgorithm}
}
\end{singlespacing}
\noindent Algorithm \ref{alg:exact}  completely circumvents MCMC based iterative sampling and only requires at most $O(n)$ flops per step. %One can now use exact sampling to generate posterior distributions for $\bbeta$ and $\sigs$ or simply use their posterior moments for inference. It only involves one time computation of quadratic forms of NNGP precision matrices and is unsurprisingly several orders of magnitude faster. 
Although the calculations need to be replicated for every $(\phi,\alpha)$ combination, unlike the MCMC based algorithms that run serially, this step can be run in parallel. Moreover, kriging is often less sensitive to the choice of the covariance parameters so cross-validation can be done at a moderately crude resolution on the $(\phi,\alpha)$ domain. Hence, the Algorithm remains extremely fast. This incredible scalability makes the conjugate NNGP model an attractive choice for ultra high-dimensional spatial data. Although this approach philosophically departs from the true Bayesian paradigm, often inference about covariance parameters is of little interest and this hybrid cross-validation approach offers a pragmatic compromise.

\section{Illustrations}\label{sec:sim}

\subsection{Implementation}
\noindent This section details two simulation experiments and the analysis of a large remotely sensed dataset. In the analyses, we consider the candidate models labeled: \emph{Sequential} defined in \cite{nngp}; \emph{Collapsed} defined in Section~\ref{sec:nngpcol}; \emph{Response} defined in Section~\ref{sec:nngpy}, and; \emph{Conjugate} defined in Section~\ref{sec:nngpex}.

Two additional analyses are provided in the web supplement. The first, Section~\ref{sec:small}, compares full GP and NNGP model parameter estimates and predictive performance. The second, Section~\ref{sec:exp4}, moves beyond the typical geostatistical setting where $\bs$ indexes data in two-dimensions, e.g., latitude and longitude, to a more general settings where data are indexed in $N$-dimensions. Such data are common in computer experiments, where $\bs$ indexes outcomes associated with a set of values on $N$ computer model inputs. Here too, we apply a Mat\'ern covariance function. Response and Conjugate model out-of-sample predictive performance is shown to be comparable with that achieved using a local approximate Gaussian processes as implemented in the \texttt{laGP} R package \citep{lagp, gramacy2016laGP}.

Samplers were programmed in \texttt{C++} and used \texttt{openBLAS} \citep{zhang13} and Linear Algebra Package (LAPACK; \url{www.netlib.org/lapack}) for efficient matrix computations. \texttt{openBLAS} is an implementation of Basic Linear Algebra Subprograms (BLAS; \url{www.netlib.org/blas}) capable of exploiting multiple processors. Additional multiprocessor parallelization used \texttt{openMP} \citep{openmp98} to improve performance of key steps within the samplers. In particular, substantial gains were realized by distributing the calculation of NNGP precision matrix components using the \texttt{openMP} \texttt{omp for} directive. Updating these matrices is necessary for each MCMC iteration in the Sequential, Response, and Collapsed models, and for each Conjugate model cross-validation iteration. An \texttt{omp for} directive with \texttt{reduction} clause was also effectively used to evaluate quadratic function (\ref{eq:nngp_qf}) found in all models.

For the Collapsed model, SuiteSparse version 4.4.5 \citep{suitesparse} provided an interface to: fill-in minimizing algorithms, e.g., AMD \citep{amestoy04} and METIS \citep{karypis98}; CHOLMOD \citep{chen08} version 3.0.6 used for supernodal openBLAS-based Cholesky factorization to obtain $\bL$ of $\bP(\tC^{-1} + \tau^{-2} \bI)\bP^\top$, and solvers for sparse triangular systems. Also see the text by \cite{davis06}.

For each analysis using the Collapsed model, nine fill-in algorithms were considered \cite[for details see][pages 4 and 16, respectively]{chen08, cholmod} for formation of the permutation matrix $\bP$. Assessment of the various fill-in algorithms is based on the resulting pattern of non-zero matrix elements. This is important for our setting because the initial pattern of the NNGP precision matrix is determined by the neighbor set and, hence, discovery of an \emph{optimal} permutation matrix need only be done once prior to sampling.

Implementing NNGP models requires a neighbor set for each observed location. For a given location $\bs_i$, a brute force approach to finding the neighbor set calculates Euclidean distances to $\bs_1$, $\bs_2$ and $\bs_{i-1}$, sorts these distances while keeping track of locations' indexes, then selects the $m$ minimum distance neighbors. This brute force approach is computationally demanding. Subsequent analyses use a relatively simple to implement fast nearest neighbor search algorithm proposed by \citet{ra93} that provides substantial efficiency gains over the brute force search (see supplemental material for details). 

All subsequent analyses were conducted on a Linux workstation with two 18-core Intel processors and 512 GB of memory. Unless otherwise noted, posterior inference used the last $1\times10^4$ iterations from each of three chains of $2.5\times10^4$ iterations. Chains run for a given model were initiated at different values and each chain was given a unique random number generator seed. Following \citet{nngp}, all models were fit using $m$=15 neighbors unless noted otherwise.

\emph{Upon publication, code and data needed to reproduce the analyses will be provided in the JCGS web supplement. While under review, the code and data are available at \url{http://blue.for.msu.edu/data/JCGS-code-data.tar.gz}.}

\subsection{Experiment \#1}
\noindent The aim of this experiment was to assess NNGP model run time. To achieve this, we selected data subsets for a range of $n$ from the TIU dataset described in Sections~\ref{sec:intro} and \ref{tanana}. The posited model follows (\ref{eq:likewy}) and includes an intercept and slope regression coefficients, and an exponential covariance function with parameters $\sigma^2$, $\phi$, and residual variance $\tau^2$. A ``flat'' improper prior distribution was assigned to each regression coefficient, $\beta$'s, which places equal weight on all possible values of the parameter. The variance components $\tau^2$ and $\sigma^2$ were assigned inverse-Gamma $IG(2,10)$ priors, and a uniform $U(0.1,10)$ prior for the decay parameter $\phi$. The support on the decay corresponds to an effective spatial range (i.e., the distance where the spatial correlation is 0.05) between 0.3 to 30 km (see Section~\ref{tanana} for specifics on the TIU domain and dataset).

\begin{figure}[!ht]
\begin{center}
	\subfigure[]{\includegraphics[width=8cm]{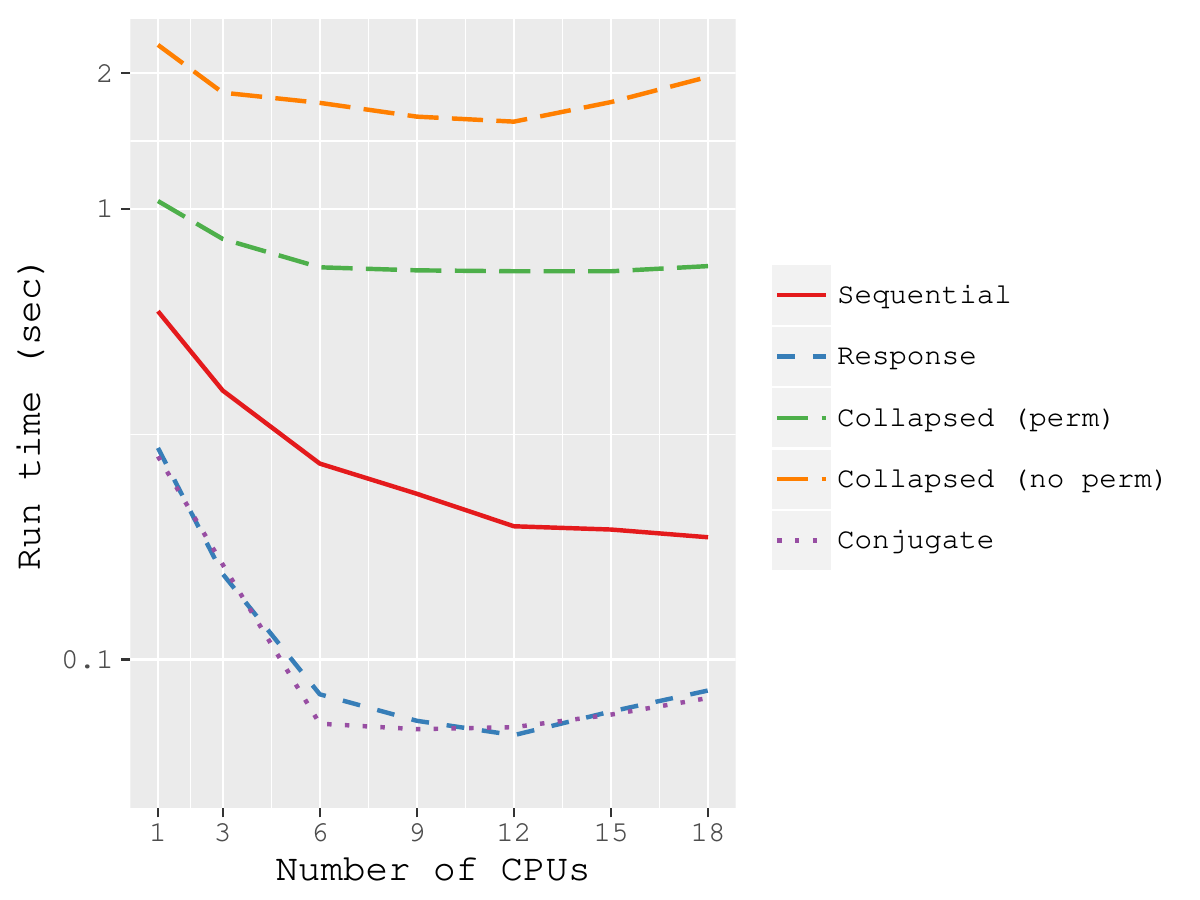}\label{cpu-timing}}
        \subfigure[]{\includegraphics[width=8cm]{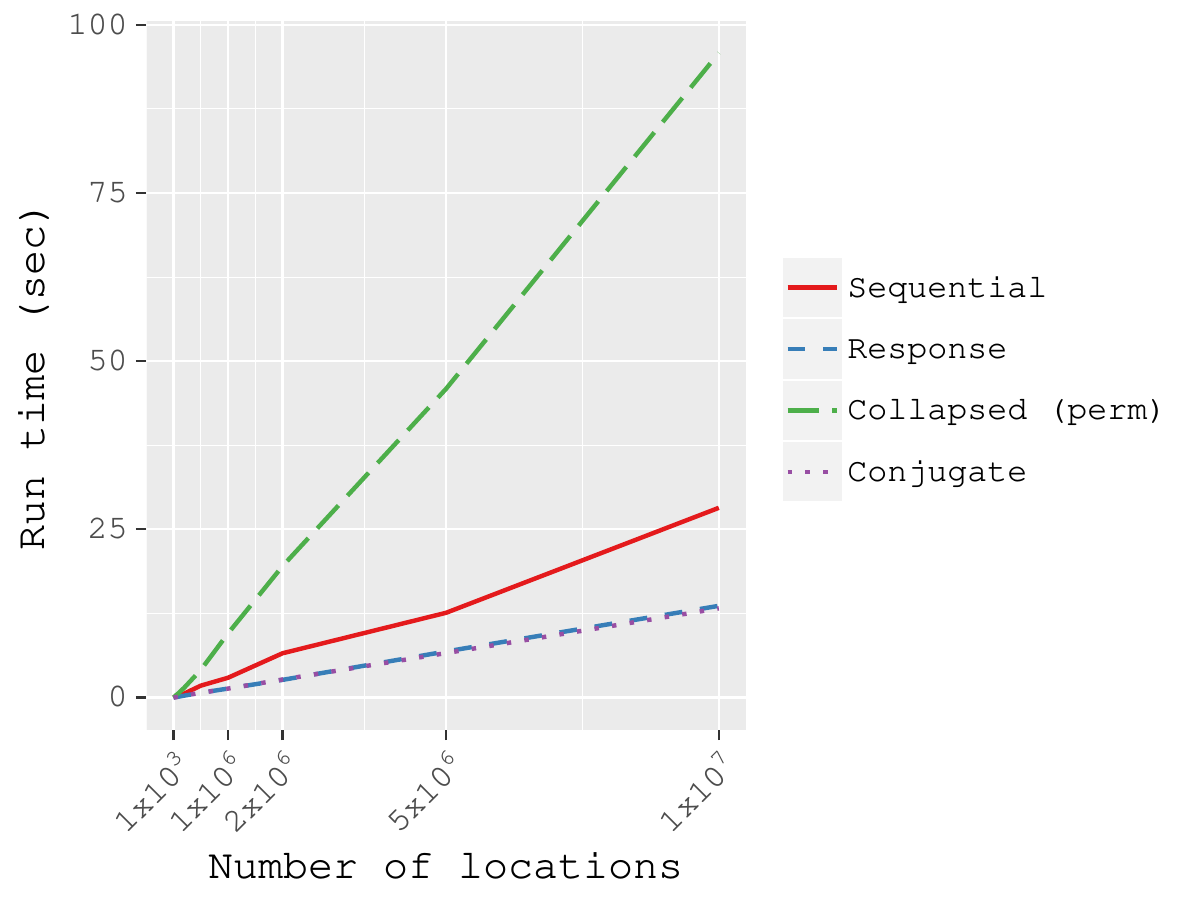}\label{n-timing}}
\end{center}
\caption{\subref{cpu-timing} Run time required for one sampler iteration using $n$=$5\times10^4$ by number of CPUs (y-axis is on the log scale). \subref{n-timing} Run time required for one sampler iteration by number of locations.}\label{timing}
\end{figure}

Figure~\ref{cpu-timing} shows run time for a dataset of $n$=$5\times 10^4$ and number of CPUs used to complete one MCMC iteration (not including the initial nearest neighbor set search time, which is common across models). Two versions of the Collapsed model are shown, one assumes the permutation matrix $\bP$ is diagonal (labeled \emph{no perm}) and the other allows CHOLMOD to select an approximately optimal permutation matrix (labeled \emph{perm}). Here, and in other experiments, using a fill-in reducing permutation matrix provides substantial time efficiency gains. The Response model provides full posterior inference on all parameters, with the exception of $\bw$, and dramatically faster run time compared to the Collapsed model. Inference for the Conjugate model, including point estimates of $\bbeta$ and $\sigma^2$, requires about the same amount of time as one Response model MCMC iteration. Explicitly updating $\bw$ is relatively slow; hence, the Sequential model's computing time falls somewhere between that of the Collapsed and Response models.

For all models, Figure~\ref{cpu-timing} show marginal improvement in run time beyond $\sim$6 CPUs and negligible improvement beyond $\sim$12 CPUs. We attribute the slight increase in run time beyond $\sim$12 CPU seen in some models to communication overhead. Run time is actual execution time, or ``wall clock time'' of the specified number of MCMC iterations. Points of diminishing return on number of CPUs used will change with $n$; however, exploratory analysis across the range of $n$ considered here suggested 12 CPUs is the bound for substantial gains (clearly this also depends on computing environment and programming decisions).

Figure~\ref{n-timing} shows time required to execute one sampler iteration by $n$. The Response and Conjugate models deliver inference across $n$ in $\sim$1/3 and $\sim$1/10 the time required by the Sequential and Collapsed models, respectively. For $n$=$1\times10^7$ the run time is approximately 28, 13, 13, and 95 seconds for the Sequential, Response, Conjugate, and Collapsed, respectively. 

\subsection{Experiment \#2}\label{exp2}
\noindent This experiment compared parameters estimates and predictive performance among the NNGP models for a large dataset. Also, the potential to identify \emph{optimal} values of $\phi$ and $\alpha$ via cross-validation was assessed for the Conjugate model. We generated observations at $6\times10^4$ locations within a unit square domain from model (\ref{eq:likewy}), the $n\times n$ spatial covariance matrix $\bC$ was formed using (\ref{eq:matern}) with $\nu$ fixed at 0.5, and the mean comprised an intercept and covariate $\bx_1$ drawn from independent $N(0,1)$. Observations were then generated using the parameter values given in the column labeled \emph{True} in Table~\ref{syn-large}. Observations at $n=5\times10^4$ of these locations, selected at random, were used to estimate model parameters. Observations at the remaining $1\times10^4$ holdout locations were used to assess model predictive performance. 

\begin{figure}[!ht]
\begin{center}
	\subfigure[RMSPE]{\includegraphics[width=8cm]{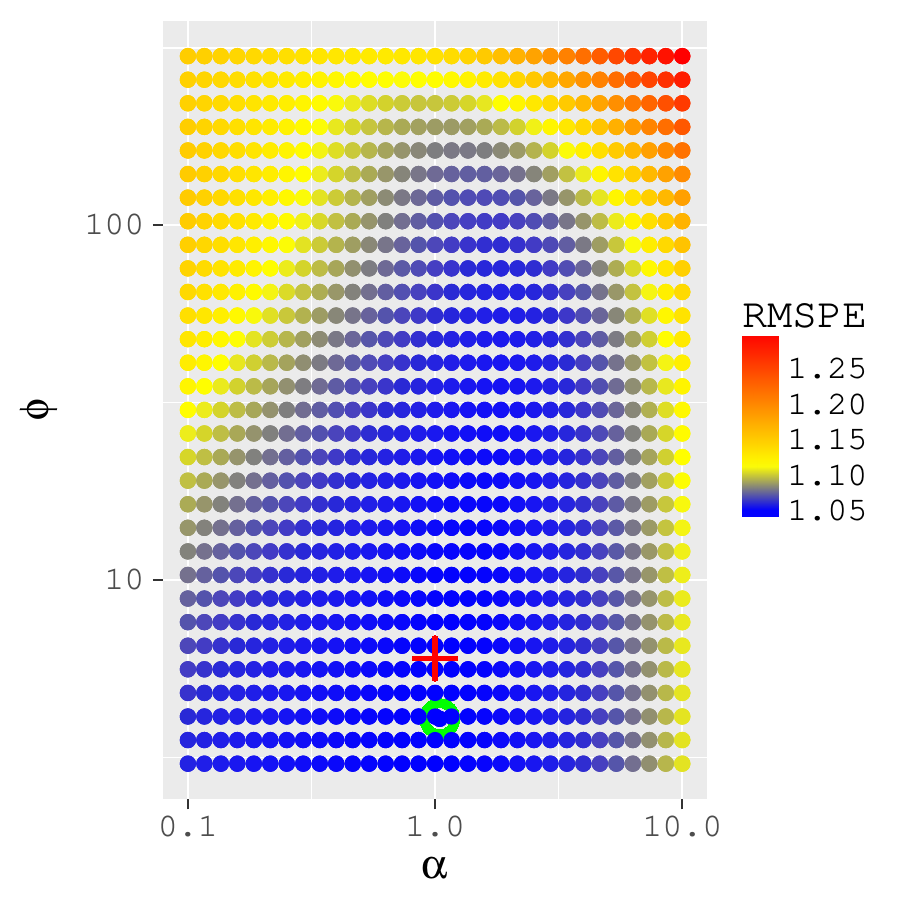}\label{syn-xval-rmspe}}
	\subfigure[CRPS]{\includegraphics[width=8cm]{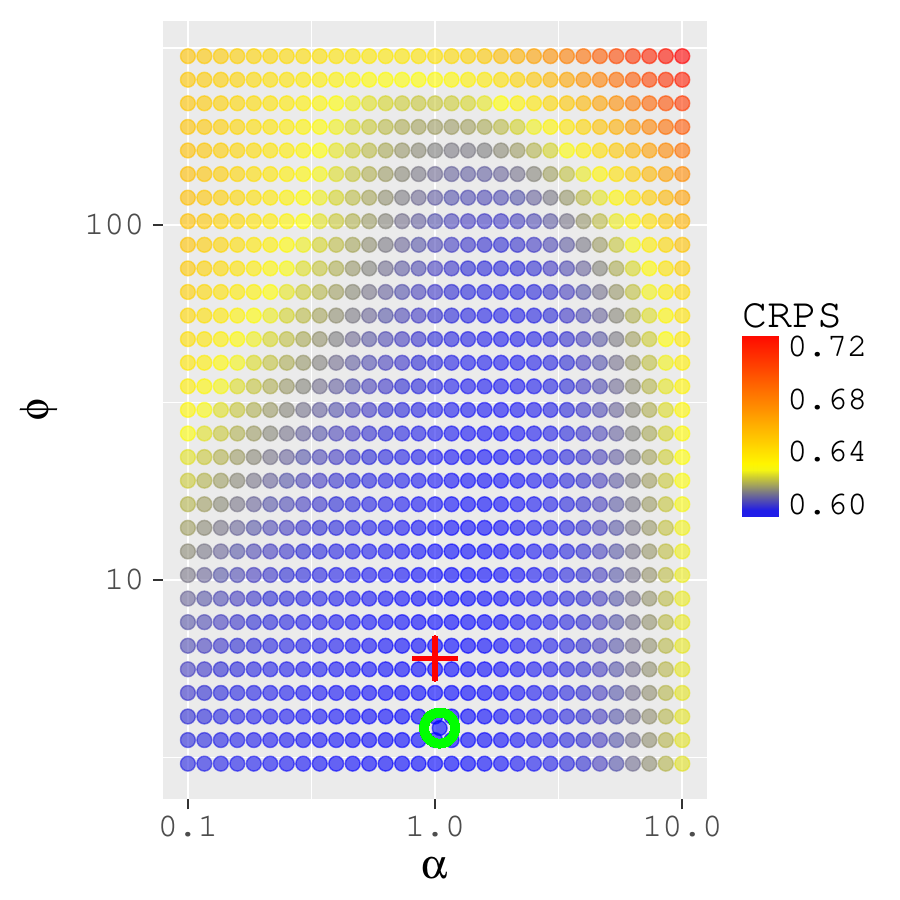}\label{syn-xval-crps}}
\end{center}
\caption{Conjugate model cross-validation results for selection of $\alpha$ and $\phi$ using the simulated dataset. Parameter combination with minimum scoring rule indicated with open circle symbol {\color{green} $\circ$} and true combination used to generate the data indicated with a plus symbol {\color{red} $+$}.}\label{syn-xval}
\end{figure}

Following Section~\ref{sec:nngpex}, five-fold cross-validation aimed at minimizing RMSPE and continuous rank probability score \citep[CRPS;][]{gneiting07} for the Conjugate model are given in Figure~\ref{syn-xval}. We observe that a broad range of $\phi$ and $\alpha$ values deliver comparable predictive performance, and minimization of RMSPE and CRPS yield approximately the same estimates of $\phi$ and $\alpha$.

In addition to RMSPE and CRPS, percent of holdout observations covered by their corresponding predictive distribution 95\% credible interval (PCI), and mean width of the predictive distributions' 95\% credible interval (PIW) were used to assess NNGP model predictive performance. Results given in Table~\ref{syn-large} show the NNGP models yield comparable parameter estimates and prediction. Here, the Conjugate model's $\phi$ and $\alpha$ were selected to minimize RMSPE (results are comparable for minimization of CRPS). 

\begin{table}[!htbp]
\begin{center}
%% \begin{sidewaystable}[!htbp]
%% \centering
\caption {Simulated dataset, parameter credible intervals $50\%\, (2.5\%,97.5\%)$ and predictive validation. Bold entries indicate where the true value is not within the 95\% credible interval.}\label{syn-large}
\scriptsize
\begin{tabular}{ccccccc}
\toprule
Parameter & True & Sequential (metrop) & Sequential (slice) & Response & Collapsed & Conjugate\\
\cmidrule{3-7}
$\beta_0$  &1&\textbf{0.64 (0.53, 0.75)} & \textbf{0.56 (0.44, 0.79)}& \textbf{0.84 (0.70, 0.99)} & 1.10 (0.51, 1.79) &  0.84\\ 
$\beta_1$  &5&5.00 (5.00, 5.01) & 5.00 (5.00, 5.01) & 5.01 (5.00, 5.01) & 5.00 (5.00, 5.01) & 5.01\\ 
$\sigma^2$  &1&\textbf{1.95 (1.44, 2.21)} & \textbf{1.68 (1.11, 2.19)}& 1.03 (0.91, 1.21) & \textbf{1.69 (1.16, 2.24)} & 0.98\\ 
$\tau^2$  &1&1.00 (0.98, 1.01) & 1.00 (0.98, 1.01)& 1.00 (0.98, 1.01) & 1.00 (0.98, 1.01) & 1.02\\ 
$\phi$  &6&\textbf{3.39 (3.03, 4.54)} & 3.98 (3.04, 6.05)& 6.26 (4.88, 7.78) & \textbf{3.95 (3.01, 5.83)} & 4.05\\ 
\midrule
CRPS  &&0.59&0.59& 0.6&0.59&0.59\\
RMSPE  &&1.04&1.04& 1.05&1.04&1.05\\
95\% PIC  &&93.13&92.63& 93.08&92.77&94.94\\
95\% PIW  &&3.87&3.85& 3.93&3.84&4.11\\
\bottomrule
\end{tabular} 
%% \end{sidewaystable}
\end{center}
\end{table}

Candidate models' Gelman-Rubin \citep{gelman92} potential scale reduction factor figures and MCMC chain trace plots are given in Figures \ref{syn-large-gr} - \ref{syn-large-col-chains} of the web supplement. These figures show the Response and Collapsed models provide faster chain convergence for the intercept and spatial covariance parameters compared to Sequential model. Additional analysis in Section~\ref{sec:small} of the web supplement reveal that for a smaller dataset generated using the same model, the Sequential model parameter posteriors do not match well that of the full GP. 

\subsection{Tanana Inventory Unit forest canopy height}\label{tanana}

Our goal is to create a high-resolution forest canopy height data product, {with accompanying uncertainty estimates for prediction and spatial correlation parameters}, for the US Forest Service Tanana Inventory Unit (TIU) that covers a large portion of Interior Alaska using a sparse sample of LiDAR data from NASA Goddard's LiDAR, Hyperspectral, and Thermal (G-LiHT) Airborne Imager \citep{cook2013}.

\noindent For remote forested regions, combining sparse airborne LiDAR data with a sparse network of forest inventory data provides a cost-effective means to deliver predictive maps of forest canopy height. In this study, LiDAR data were acquired across the US Forest Service Tanana Inventory Unit (TIU) in Interior Alaska, approximately 140,000 km$^2$, using the NASA Goddard's LiDAR, Hyperspectral, and Thermal (G-LiHT) Airborne Imager \citep{cook2013}. The G-LiHT instrument package simultaneously acquires data from a suite of remote sensing instruments to collect complementary information on forest structure (LiDAR), vegetation composition (hyperspectral), and forest health (hyperspectral and thermal). 

\begin{figure}[!ht]
\begin{center}
	\subfigure[]{\includegraphics[width=8cm]{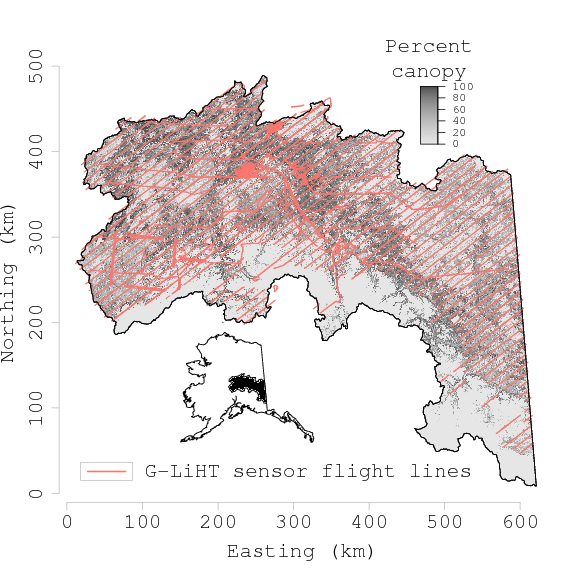}\label{han}}
	\subfigure[]{\includegraphics[width=8cm]{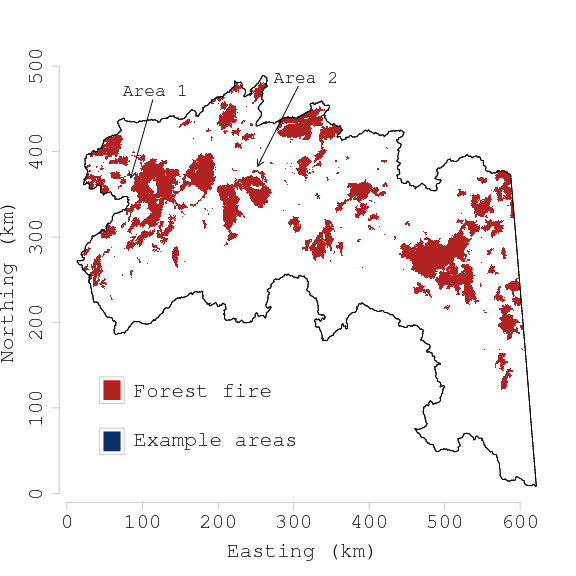}\label{fire}}
\end{center}
\caption{TIU, Alaska, study region. \subref{han} G-LiHT flight lines where canopy height was measured at $5\times10^6$ locations and percent tree cover predictor variable. \subref{fire} Occurrence of forest fire within the past 20 years predictor variable and two example areas for prediction illustration.}
\end{figure}

Here, we consider G-LiHT LiDAR data collected during a 2014 TIU flight campaign. The campaign collected a systematic sample covering $\sim$8\% of the TIU, with 78 parallel flight lines spaced $\sim$9 km apart, Figure~\ref{han}, along with incidental measurements to-and-from the transects. The nominal flying altitude of data collection in the TIU was 335 m above ground level, resulting in a sample swath width of $\sim$180 m (30$^{\circ}$ field of view) and sample density of 3 laser pulses m$^2$. Point cloud data were classified and used to generate bare earth elevation and canopy height models at 1 m ground sample distance, as described in \cite{cook2013}. G-LiHT point cloud data and derived products are available online at \url{http://gliht.gsfc.nasa.gov}. The data was processed following methods in \cite{cook2013}, such that 28,751,400 LiDAR-based estimates of forest canopy height were available on a 15$\times$15 m grid along the flight lines. Each grid cell yielded an estimate of canopy height calculated as the height below which 95\% of the pulse data was recorded. The subsequent analysis uses a random sample of $5.025\times10^6$ observations from the larger LiDAR dataset.

Two predictors that completely cover the TIU were considered. First, a Landsat derived percent tree cover data product developed by \cite{hansen13}, shown as the gray scale surface in Figure~\ref{han}. This product provides percent tree cover estimates for peak growing season in 2010 (most recent year available) and was created using a regression tree model applied to Landsat 7 ETM+ annual composites. These data are provided by the United States Geological Survey (USGS) on an approximate 30 m grid covering the entire globe \citep{hansen13}. Second, the perimeters of past fire events from 1947-2014 were obtained from the Alaska Interagency Coordination Center Alaska fire history data product \citep{AICC}. Forest recovery/regrowth following fire is very slow in Interior Alaska. Hence we discretized the fire history data to 1 if the fire occurred within the past 20 years and 0 otherwise, Figure~\ref{fire}.

We explored the relationship between canopy height, tree cover and fire history using a non-spatial regression model and NNGP Response, Collapsed, and Conjugate models. We did not consider the Sequential model here because of the convergence issues seen in the preceding experiments. Exploratory analysis using the non-spatial regression suggested both predictors explain a substantial portion of variability in canopy height (Table~\ref{ak-large}), with a positive association between canopy height and tree cover (TC) and negative association between canopy height and recent fire occurrence (Fire). These results are consistent with our understanding of the TIU forest system. The tree cover variable captures forest canopy sparseness---with sparser canopies resulting in LiDAR height percentiles shifted toward the ground. Recently burned areas are typically replaced with regenerating, shorter stature, forests.% A variogram analysis using the non-spatial model's residuals suggested a substantial portion of the residual variance can be attributed to spatial association and the spatial range of $\sim$5 km agrees with visual assessment of forest patch structure across much of the TIU (Figure~\ref{ak-vario}). These results encouraged further exploration using the NNGP space-varying intercept models developed in the preceding sections.

For all models, the intercept and slope regression parameters were given flat prior distributions. The variance components $\tau^2$ and $\sigma^2$ were assigned inverse-Gamma $IG(2,10)$ priors. We assumed an Exponential spatial correlation function with a uniform $U(0.1,10)$ prior on the decay parameter. The support on the decay corresponds to an effective spatial range between 0.3 to 30 km. Observations at $n$=$5\times10^6$ locations, selected at random, were used to estimate model parameters. Observations at the remaining $2.5\times10^4$ holdout locations were used to assess model predictive performance. 
%For the Conjugate model, selection of $\phi$ and $\alpha$  cross-validation to minimize RMSPE and CRPS. Similar to the simulated data experiment, Figure~\ref{ak-xval} shows prediction performance is fairly robust to choice of $\phi$ but sensitive to $\alpha$ values where $\tau^2$ dominates.
%
%\begin{figure}[!ht]
%\begin{center}
%	\subfigure[RMSPE]{\includegraphics[width=8cm]{ak-m15-reg-rmspe.pdf}}
%	\subfigure[CRPS]{\includegraphics[width=8cm]{ak-m15-reg-crps.pdf}}
%\end{center}
%\caption{Conjugate model cross-validation results for selection of $\alpha$ and $\phi$ using the TIU dataset. Parameter combination with minimum scoring rule indicated with open circle symbol {\color{green} $\circ$}.}\label{ak-xval}
%\end{figure}
Parameter estimates and prediction performance summaries for candidate models are given in Table~\ref{ak-large}. Results for the $m$=15 and $m$=25 models were indistinguishable, hence only $m$=15 results are presented. Here, NNGP models provide approximately the same predictive performance, and a substantial improvement over the non-spatial regression.

As suggested by Figure~\ref{n-timing}, and seen again here, the Collapsed model using a fill reducing permutation and 12 CPU requires an excessively long run time, i.e., about two weeks to generate $25\times10^3$ MCMC samples. If one is willing to forgo estimates of spatial random effects, the Response model offers greatly improved run time, i.e., about 1.5 days, and parameter and prediction inference comparable to the Collapsed model. The Conjugate model delivers the shortest run time and predictive inference comparable to the other NNGP models.

\begin{table}[!htbp]
\begin{center}
%\begin{sidewaystable}[!htbp]
%\centering
\caption {TIU dataset results. Parameter credible intervals, $50\%\, (2.5\%,97.5\%)$, predictive validation, and run time for $25\times10^3$ MCMC iterations. }\label{ak-large}
\scriptsize
\begin{tabular}{ccccc}
\toprule
& Non-spatial & & & Conjugate \\
Parameter  &regression& Response& Collapsed & minimize RMSPE\\
\cmidrule{2-5}
$\beta_0$  &-2.46 (-2.47,-2.45)&2.37 (2.31,2.42)&2.41 (2.35, 2.47)&2.51\\
$\beta_{TC}$  &0.13 (0.13, 0.13)&0.02 (0.02, 0.02)&0.02 (0.02, 0.02)&0.02\\
$\beta_{Fire}$  &-0.13(-0.14, -0.12)&0.43 (0.39, 0.48)&0.39 (0.34, 0.43)&0.35\\
$\sigma^2$  &--&17.29 (17.13, 17.41)&18.67 (18.50, 18.81)&23.21\\
$\tau^2$  &17.39 (17.37, 17.41)&1.55 (1.54, 1.55)&1.56 (1.55, 1.56)&1.21\\
%$\phi$  &--&0.0042 (0.0041, 0.0042)&0.0037 (0.0037, 0.0038)&0.001&0.0042\\
$\phi$  &--&4.15 (4.13, 4.19) &3.73 (3.70, 3.77) &3.83\\
$\alpha$ &--&--&--&0.052\\
\midrule
CRPS  &2.3&0.86&0.86&0.84\\
RMSPE  &4.19&1.72&1.73&1.71\\
95\% PIC  &93.43&94.29&94.25&94.85\\
95\% PIW  &16.27&6.58&6.56&6.73\\
\midrule
Run time (hours)& -- & 38.29 & 318.81 & 0.002\\
\bottomrule
\end{tabular} 
%\end{sidewaystable}
\end{center}
\end{table}

\begin{figure}[!ht]
\begin{center}
	\subfigure[Area 1, posterior predictive mean]{\includegraphics[width=8cm]{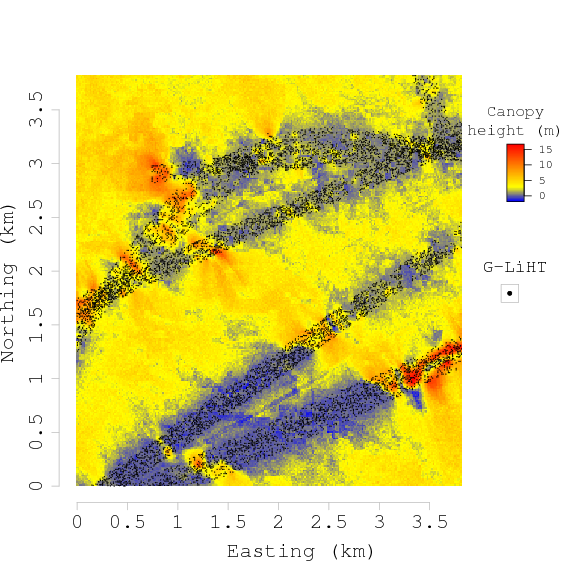}\label{aoi1mu}}
	\subfigure[Area 1, posterior predictive S.D.]{\includegraphics[width=8cm]{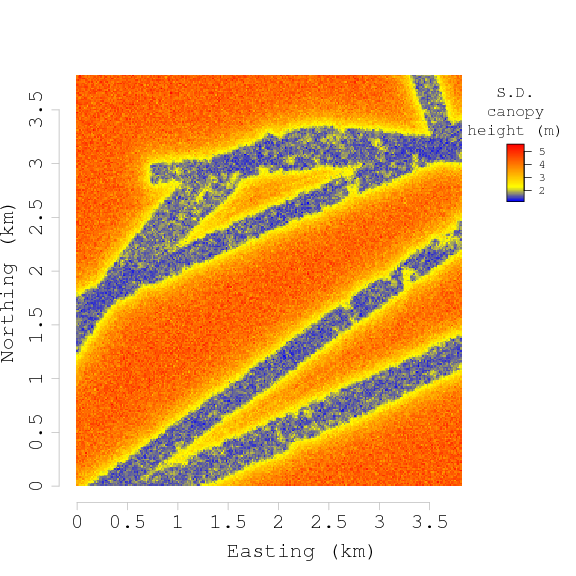}\label{aoi1sd}}\\
	\subfigure[Area 2, posterior predictive mean]{\includegraphics[width=8cm]{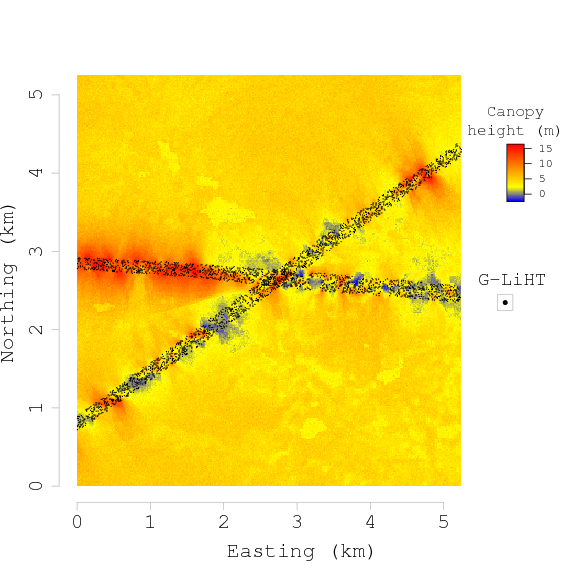}\label{aoi2mu}}
	\subfigure[Area 2, posterior predictive S.D.]{\includegraphics[width=8cm]{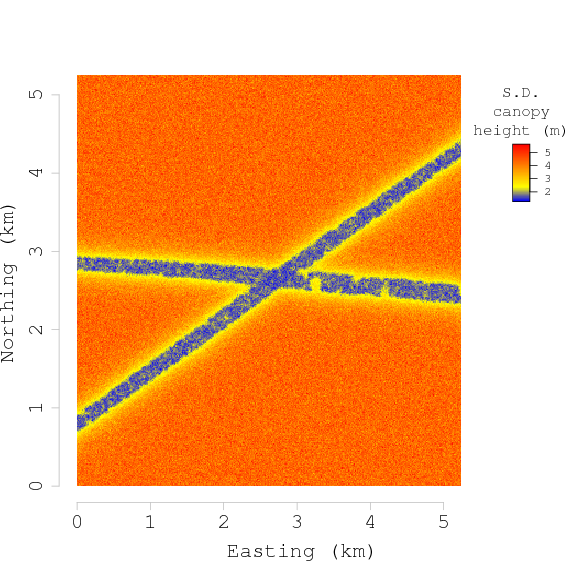}\label{aoi2sd}}
\end{center}
\caption{95th LiDAR percentile height posterior predictive distribution summary at a 30 m pixel resolution for the two example areas identified in Figure~\ref{fire}. }\label{aois}
\end{figure}

Figure~\ref{fire} identifies two example areas selected to illustrate how LiDAR and the other data inform forest canopy height prediction. As suggested by the prediction metrics in Table~\ref{ak-large}, all three NNGP models delivered nearly identical prediction map products. Figure~\ref{aois} shows the posterior predictive distribution mean and standard deviation from the Response model with $m$=15 for the two areas. Here, the left subplots identify LiDAR data locations as black points along the flight lines. The presence of strong residual spatial autocorrelation results in fine-scale prediction within, and adjacent to, the flight lines (Figures~\ref{aoi1mu}\subref{aoi2mu}) and more precise posterior predictive distributions as reflected in the standard deviation maps (Figures~\ref{aoi1sd}\subref{aoi2sd}). Predictions more than a km from the flight lines are informed primarily by tree cover and fire occurrence predictors.

The TIU forest's vertical and horizontal structure is highly heterogeneous due, in large part to topography, hydrology, and disturbance history, e.g., fire. This heterogeneity is reflected in the relatively short estimated effective range of just over 1 km (Table~\ref{ak-large}). %The discrepancy between the variogram- and model-based estimates of $\phi$ is largely due the amount of data considered. Because of memory limitations, the variogram was constructed using $2.5\times10^3$ randomly selected observations. We noticed in exploratory data analysis that as $n$ increased the estimated effective spatial range generally decreased; however, beyond $n\approx$ $2\times$10^5$ estimates stabilized to those shown in Table~\ref{ak-large}. A likely explanation is that models fit using low sample size, sparse coverage, i.e., long inter-site distance between observations, are picking up on broad landscape scale patterns in the residual spatial structure. In contrast, fine spatial resolution residual structure dominates when models are fit using large sample size with locally dense coverage. 

These results provide key input needed for planning future LiDAR campaigns to collect data to inform canopy height models. Using more informative predictor variables would certainly improve prediction across the TIU; however, few complete-coverage high spatial resolution data layers exist, other than those produced using moderate spatial resolution remote sensing products, e.g., the Landsat based tree cover predictor used here.

As seen here, high spatial resolution wall-to-wall map predictions can be achieved with sufficient LiDAR coverage and use of fine-scale residual spatial structure. The G-LiHT LiDAR data---spatially dense along the 180 m swath widths---could better inform canopy height prediction across the TIU if it covered a larger swath width. This could be accomplished by increasing the flight altitude. While a higher nominal flying altitude will increase the swath width, it will also decrease the spatial density of LiDAR observations. Our results suggest that LiDAR density is less important than coverage width, given models were fit using only $\sim$17\% ($5\times10^6$/28,751,400) of available data and even then it appears we had ample information to inform prediction within flight lines. This observation, has implications for the other LiDAR collection campaigns, e.g., ICESat-2 \citep{abdalati2010, ICESAT2} and Global Ecosystem Dynamics Investigation LiDAR \citep{GEDI2014}, when they choose between pules density and swath width.

\section{Summary}\label{sec:summary}
\noindent Our aim has been to propose alternate formulations and derivatives of Bayesian NNGP models developed by \cite{nngp} to substantially improve computational efficiency for fully process-based inference. These improvements make it feasible to bring a rich set of hierarchical spatial Gaussian process models to bear on data intensive analyses such as the TIU forest canopy mapping effort. Analysis of simulated data shows that compared with the Sequential specification of \cite{nngp}, the Response and Collapsed models offer improved MCMC chain behavior for the intercept and spatial covariance parameters. If full inference about the spatial random effects is of interest, then the Response or Conjugate models are not appropriate. So while the Collapsed model can be computationally intensive, depending on the burden imposed by the sparse Cholesky decomposition, it is the only fully Bayesian alternative to the sequential Gibbs sampler developed in \cite{nngp} and should generally be selected over the latter due to its significantly improved chain convergence. Furthermore, recent work by \cite{katzguin} shows that the collapsed model provides a better approximation of the full GP than the Response model in the sense of Kullback-Leibler divergence from the full GP model. If model parameter estimation and/or spatial interpolation of the response is the primary objective, the Response model offers substantial computational gains over the Collapsed model. Finally, relative to the other NNGP models, the Conjugate model delivers massive gains in computational efficiency and seemingly uncompromised predictive inference, but requires specification of the models' spatial decay and $\alpha$ parameters. However, as demonstrated in the simulation and TIU analyses, these parameters can be effectively selected via cross-validation. The response and conjugate NNGP models are available for public use in the \texttt{spNNGP} package \citep{spnngp} in R. 

The Response model emerges a viable option for obtaining full Bayesian inference about spatial covariance parameters and prediction units. A fully Bayesian kriging model capable of handling $5\times10^6$ observations on standard computing architectures is an exciting advancement and opens the door to using a rich set of process models to tackle complex problems in big data settings. For example, the Response and Collapsed NNGP models can seamlessly replace Gaussian Processes within multivariate, space-varying coefficients, and space-time settings \citep[see, e.g.,][]{nngp, dnngp, wiresnngp}. The Conjugate model provides a new tool for delivering fast interpolation with few inferential concessions. Extension of the Conjugate model to some of the more complex hierarchical frameworks noted above provides an additional avenue for development.

The TIU analysis shows the advantage of embedding the NNGP as a sparsity-inducing prior within a hierarchical modeling framework. The proposed NNGP specifications yield complete coverage forest canopy height prediction maps with associated uncertainty estimates using sparsely sampled but locally dense $n=5\times10^6$ LiDAR data. The resulting data product is the first statistically robust map of forest canopy for the TIU. Insight into residual spatial dependence will help guide planning for upcoming LiDAR data collection campaigns at global and local scales to improve prediction by leveraging information in more optimally located canopy height observations.

There remains much to be explored in NNGP models. Recent investigations by \cite{guinness16} suggest that the Kullback-Leibler divergence between full Gaussian process likelihoods and Vecchia-type nearest neighbor approximations can be sensitive to topological ordering. Our preliminary explorations seem to suggest that while the Kullback-Leibler divergence from the truth may be affected, substantive inference in the form of parameter estimates and predictive performance (based upon root-mean-square-predictions) are very robust. Nevertheless, we are currently conducting further investigations with the ordering suggested by \cite{guinness16} and intend to report on our findings in a subsequent work. Another pertinent matter concerns the performance of NNGP models for nonstationary processes. Naive implementations using neighbor selection based on simple Euclidean metrics may not be desirable. Here, the dynamic neighbor-finding algorithms proposed by \cite{dnngp} in spatiotemporal contexts may offer a better starting point than finding suitable metrics to choose neighbors. Still, work needs to be done in developing and analyzing analogous algorithms for nonstationary processes. Finally, there is scope to explore NNGP models for high-dimensional multivariate outcomes using spatial factor models \cite{taylor18} or Graphical Gaussian models and assessing their efficiency for highly complex multivariate spatial datasets.        

\section*{Acknowledgments}
\noindent Finley was supported by National Science Foundation (NSF) DMS-1513481, EF-1137309, EF-1241874, and EF-1253225. Cook, Morton, and Finley were supported by NASA Carbon Monitoring System grants. Banerjee was supported by NSF DMS-1513654, NSF IIS-1562303 and NIH/NIEHS R01-ES027027.

%\bibliographystyle{asa}
%\bibliographystyle{apalike}
%\bibliography{FDB16bib}

\clearpage 
\renewcommand{\thesection}{S\arabic{section}}
\renewcommand{\thefigure}{S\arabic{figure}}
\renewcommand{\thetable}{S\arabic{table}}

\setcounter{table}{0}    
\setcounter{figure}{0}    
\setcounter{section}{0}    

\begin{center}
{\large Web supplement for \emph{Efficient algorithms for Bayesian Nearest Neighbor Gaussian Processes}}
\end{center}

\section{Fast nearest neighbor search}\label{sec:nnsearch}
\begin{figure}[!ht]
  \begin{center}
    \includegraphics[width=16cm]{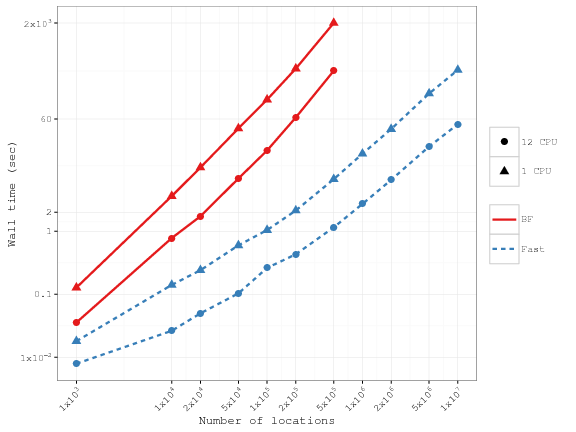}
  \end{center}
  \caption{Wall time required for neighbor set search using a brute force (BF) and fast search Algorithm by $n$ and number of CPUs.}\label{search-time}
\end{figure}

Construction of NNGP models require a neighbor set for each observed location. Identifying these sets is trivial when $n$ is small. For a given location $\bs_i$, a brute force approach to finding the neighbor set calculates the Euclidean distances to $\bs_1$, $\bs_2$ and $\bs_{i-1}$, sorts these distances while keeping track of locations' indexes, then selects the $m$ minimum distance neighbors. Figure~\ref{search-time} shows the time required to perform this \emph{brute force} approach for a range of $n$. Here, we can see that when $n$ is larger than $\sim$5x$10^4$ the brute force approach becomes prohibitively slow. Distributing this brute force search over multiple CPUs, e.g., 12 CPUs also shown in Figure~\ref{search-time}, does improve performance; however, $n$ larger than $\sim$5x$10^5$ is again too slow. Developing data structures and associated Algorithms for efficient nearest neighbor searches is a major focus in computer science and engineering, see, e.g., \citet{buhlmann16} Ch. 7. Given the size of datasets considered in subsequent analyses, i.e., $<$1x$10^7$, we chose a relatively simple to implement fast nearest neighbor search Algorithm proposed by \citet{ra93} that provides substantial efficiency gains over the brute force search as shown in Figure~\ref{search-time} (labeled \emph{fast}). We note that future work could look into modifying more sophisticated structures and associated search Algorithms such as binary search trees \citep{cormen09} to deliver the nearest neighbor set more efficiently given the NNGP ordering constraint.

\clearpage
\section{Experiment \#2}\label{sec:datasubset}
\begin{figure}[!ht]
  \begin{center}
    \includegraphics[width=12cm]{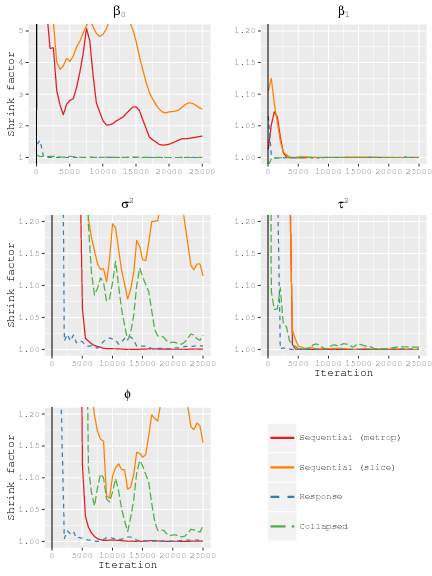}
  \end{center}
  \caption{Simulated dataset, Gelman-Rubin convergence diagnostics plots for candidate model parameters. Results consider both a Metropolis Hastings (metrop) and Slice sampler (slice) update of the Response models' covariance parameters.}\label{syn-large-gr}
\end{figure}

\begin{figure}[!ht]
  \begin{center}
    \includegraphics[width=12cm]{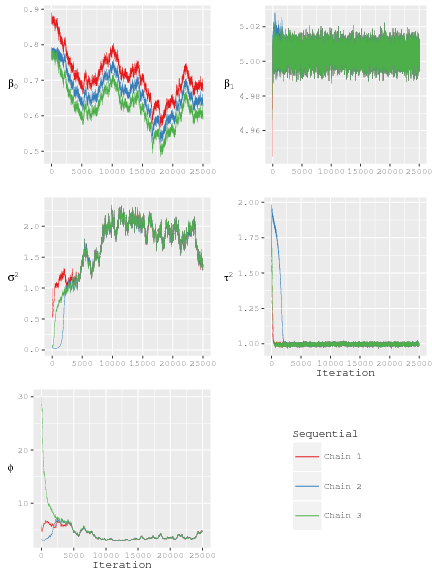}
  \end{center}
  \caption{Simulated dataset, Sequential model MCMC chain trace plots.}\label{syn-large-seq-chains}
\end{figure}

\begin{figure}[!ht]
  \begin{center}
    \includegraphics[width=12cm]{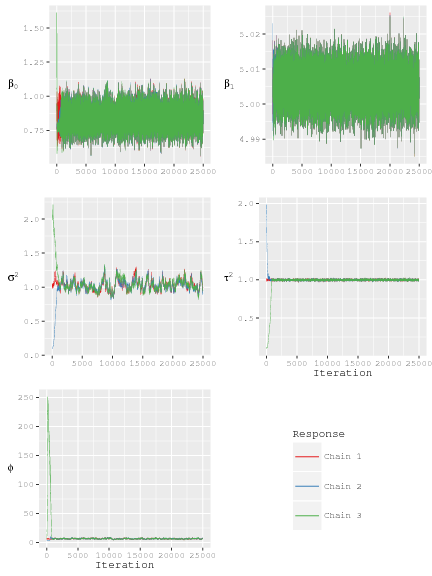}
  \end{center}
  \caption{Simulated dataset, Response model MCMC chain trace plots.}\label{syn-large-resp-chains}
\end{figure}

\begin{figure}[!ht]
  \begin{center}
    \includegraphics[width=12cm]{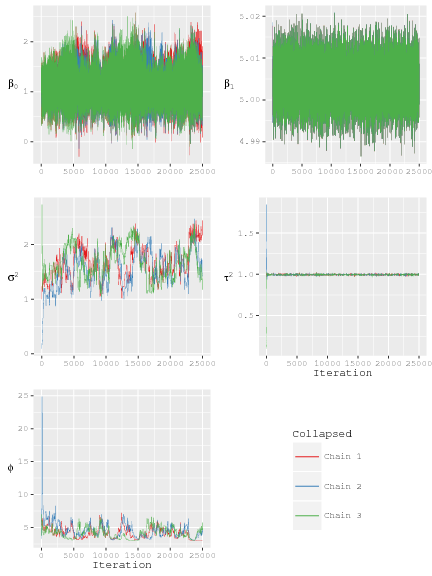}
  \end{center}
  \caption{Simulated dataset, Collapsed model MCMC chain trace plots.}\label{syn-large-col-chains}
\end{figure}

\clearpage
\section{Experiment \# 3}\label{sec:small}
Simulated data were generated from the model specified in Experiment \#2 detailed in Section 3.3 with the exception that $n=1500$. Observations at $n=1000$ of these locations, selected at random, were used to estimate model parameters. Observations at the remaining $500$ holdout locations were used to assess model predictive performance. Specifically, given the holdout observations and model posterior predictive distribution samples, predictive performance was summarized using: 1) mean continuous rank probability score (CRPS), which is a strictly proper scoring rule that quantifies the fit of the entire predictive distribution (i.e., for a normal distribution, the mean and the variance) to the data \citep{gneiting07}; 2) root mean-square prediction error (RMSPE) between observed values and means of the predictive distributions; 3) percent of observations covered by their corresponding predictive distribution 95\% credible interval (PCI), and; mean width of the predictive distributions' 95\% credible interval (PIW). 

The experiment sample size was kept purposely small so we could compare NNGP model performance to that of a full GP model. Full GP model parameter estimates and predictions were obtained using the spBayes R package \texttt{spLM} function \citep{finley15}. For all models, the prior distribution on regression coefficients $\beta_0$ and $\beta_1$ were assumed to be \emph{flat} and variance parameters $\sigma^2$ and $\tau^2$ followed an inverse-gamma $IG(2, 1)$. For the full GP, Sequential, Response, and Collapsed models the spatial decay parameter $\phi$ followed a uniform $U(3, 300)$. This prior support for $\phi$ assumes the effective range is between 0.01 and 1 distance units (where the effective range is defined as the distance at which the correlation equals 0.05). As detailed in Section 1.4, the Conjugate model $\phi$ and $\alpha$ were selected to minimize RMSPE using a 5-fold cross-validation.

\begin{figure}[!ht]
  \begin{center}
    \includegraphics[width=12cm]{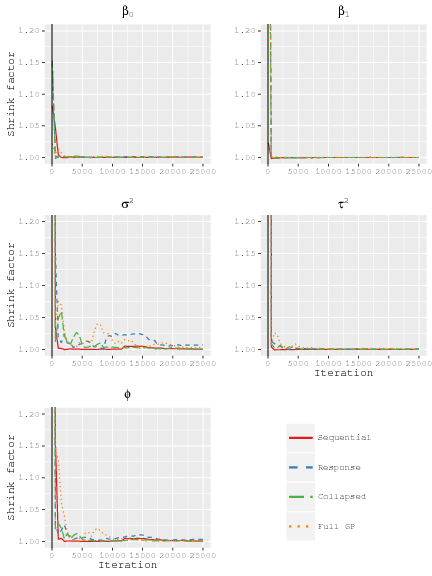}
  \end{center}
  \caption{Simulated dataset, Gelman-Rubin convergence diagnostic plots for NNGP and full GP model parameters.}\label{syn-small-gr}
\end{figure}

\begin{figure}[!ht]
  \begin{center}
    \includegraphics[width=12cm]{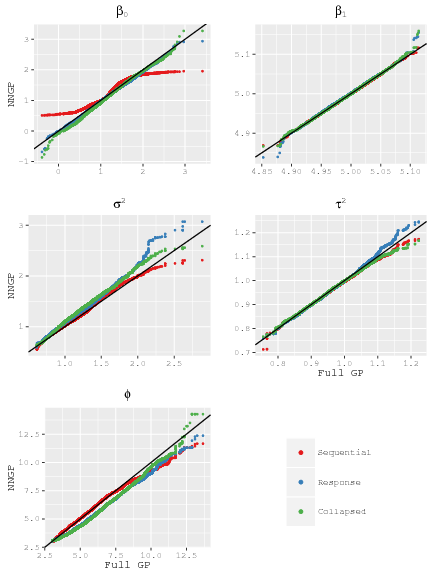}
  \end{center}
  \caption{Simulated dataset, quantile-quantile plots of NNGP versus full GP model parameter posterior distributions.}\label{syn-small-qq}
\end{figure}

\begin{figure}[!ht]
  \begin{center}
    \includegraphics[width=12cm]{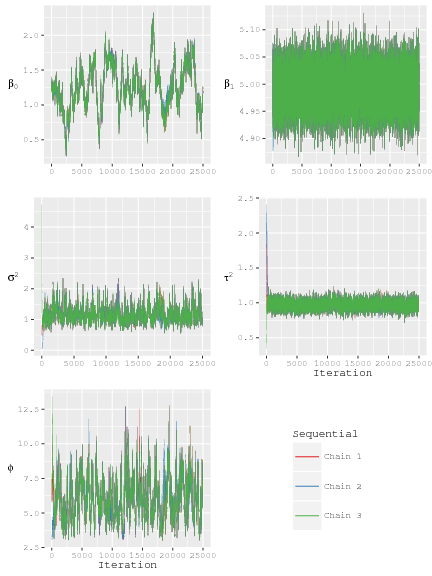}
  \end{center}
  \caption{Simulated dataset,  Sequential $m$=15 model MCMC chain trace plots.}\label{syn-small-seq-chains}
\end{figure}

\begin{figure}[!ht]
  \begin{center}
    \includegraphics[width=12cm]{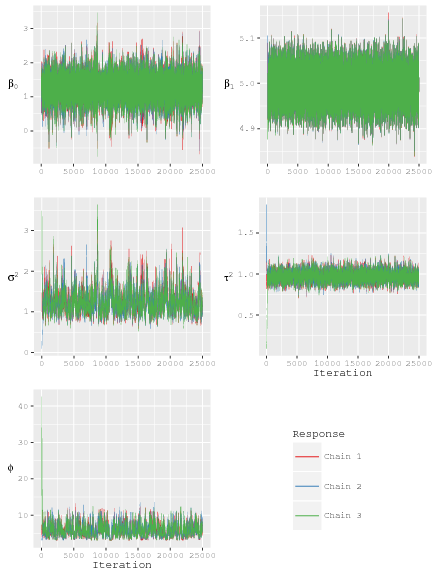}
  \end{center}
  \caption{Simulated dataset, Response $m$=15 model MCMC chain trace plots.}\label{syn-small-resp-chains}
\end{figure}

\begin{figure}[!ht]
  \begin{center}
    \includegraphics[width=12cm]{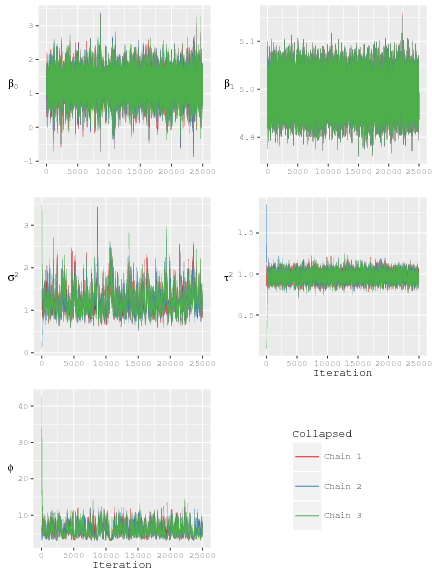}
  \end{center}
  \caption{Simulated dataset, Collapsed $m$=15 model MCMC chain trace plots.}\label{syn-small-col-chains}
\end{figure}

The aim of this experiment was to assess samplers' convergence characteristics and determine to what extent NNGP model inference compares to the full GP model. The Gelman-Rubin \citep{gelman92} potential scale reduction factor, Figure~\ref{syn-small-gr}, and visual inspection of MCMC chain trace plots Figures~\ref{syn-small-seq-chains}-\ref{syn-small-col-chains} showed adequate convergence and mixing within a thousand MCMC iterations for all models. Parameter estimates and predictive performance metrics in Table~\ref{syn-small} suggest that for these data, NNGP models do indeed deliver inference comparable to the full GP model. Results from more extensive simulation experiments conducted by \citet{nngp} and \citet{dnngp} comparing the Sequential and full GP models corroborate our findings. Quantile-quantile (Q-Q) plots in Figure~\ref{syn-small-qq} provide a more detailed comparison between the posterior distributions generated using the NNGP and full GP models. Here, the Sequential model's $\beta_0$ shows the most striking departure from the full GP estimates---with tails substantially shorter than those of the full GP. There were some other differences in the tails of the NNGP posterior distributions compared to the full GP model estimates; however, these were relatively minor. Increasing the nearest neighbor set size from 15 to 25 did not have a substantial impact on the NNGP model posterior distributions or predictive inference (Table~\ref{syn-small}).

\begin{table}[!htbp]
\begin{center}
%\begin{sidewaystable}[!htbp]
\centering
\caption {Simulated dataset, model parameter estimates $50\%\, (2.5\%,97.5\%)$ and predictive validation.}\label{syn-small}
\scriptsize
\begin{tabular}{ccccccc}
  \toprule
  Parameter & True & Full GP &Sequential (m=15) & Response (m=15)& Collapsed (m=15)& Conjugate (m=15)\\
\midrule
$\beta_0$  &1&1.28 (0.65, 1.99)&1.25 (0.74, 1.92) & 1.25 (0.56, 1.91) & 1.24 (0.50, 1.93) &1.28\\ 
$\beta_1$  &5&4.99 (4.93, 5.05)&4.99 (4.93, 5.06) & 4.99 (4.93, 5.06) & 4.99 (4.93, 5.06) &4.99\\ 
$\sigma^2$  &1&1.11 (0.77, 1.85)&1.10 (0.76, 1.79) & 1.20 (0.80, 1.99) & 1.18 (0.78, 1.95) &1.17\\ 
$\tau^2$  &1&0.96 (0.86, 1.10)&0.97 (0.85, 1.09) & 0.97 (0.85, 1.10) & 0.97 (0.85, 1.10) &0.9\\ 
$\phi$  &6&6.06 (3.34, 9.71)&6.08 (3.40, 9.61) & 5.45 (3.20, 9.31) & 5.53 (3.18, 9.50)&7.89\\ 
\midrule
CRPS  &     &0.65   &0.66  &0.66  &0.66  &0.65\\
RMSPE  &    &1.15   &1.15  &1.16  &1.15  &1.15\\
95\% PIC  & &92.8   &93.8  &92.4  &92    &--\\
95\% PIW  & &4.17   &4.32  &4.18  &4.09  &--\\
\cmidrule{1-7}
& & Sequential (m=25) & Response (m=25)& Collapsed (m=25)& Conjugate (m=25)&\\
\cmidrule{1-6}
$\beta_0$  &1& 1.27 (0.78, 1.94) & 1.27 (0.58, 1.95) & 1.27 (0.58, 1.94) &1.28&\\ 
$\beta_1$  &5& 4.99 (4.93, 5.06) & 4.99 (4.93, 5.06) & 4.99 (4.93, 5.06) &4.99&\\ 
$\sigma^2$  &1&1.05 (0.74, 1.71) & 1.08 (0.74, 1.79) & 1.12 (0.75, 1.83) &1.1&\\ 
$\tau^2$  &1&0.97 (0.85, 1.09) & 0.98 (0.86, 1.12) & 0.97 (0.85, 1.10) &0.92&\\ 
$\phi$  &6&6.42 (3.74, 9.86) & 5.87 (3.26, 9.51) & 5.92 (3.30, 9.73) &7.89&\\ 
\cmidrule{1-6}
CRPS  &     &0.66  &0.66  &0.66  &0.65 &\\
RMSPE  &    &1.15  &1.16  &1.15  &1.14 &\\
95\% PIC  & &94    &92.4  &92.4  &--   &\\
95\% PIW  & &4.32  &4.18  &4.10   &--  &\\
\cmidrule{1-6}
\end{tabular} 
%\end{sidewaystable}
\end{center}
\end{table}

%% \begin{figure}[!ht]
%% \begin{center}
%%   \includegraphics[width=8cm]{non-sp-resid-variog.pdf}
%% \end{center}
%% \caption {Variogram of TIU non-spatial model residuals with an Exponential variogram model fit curve. Lower and upper horizontal lines correspond to $\tau^2$ and $\sigma^2$ parameter estimates, respectively, and vertical line corresponds to the distance at which the correlation drops to 0.05.}\label{ak-vario}
%% \end{figure}

%% \begin{figure}[!ht]
%%   \begin{center}
%%     \includegraphics[width=12cm]{syn-large-m15-gr.png}
%%   \end{center}
%%   \caption{Simulated dataset \#2, MCMC chain trace plots for NNGP and Full GP model parameters.}\label{syn-large-chains}
%% \end{figure}

\clearpage
\section{Experiment \#4}\label{sec:exp4}
\noindent While our focus to this point has been on inference for processes existing within the typical geostatistical setting where $\bs$ indexes data in two-dimensions, e.g., latitude and longitude, there is also interest in more general settings where data are indexed in $N$-dimensions. Such data are common in computer experiments, where $\bs$ indexes outcomes associated with a set of values on $N$ computer model inputs, see., e.g., \cite{santner03} for foundations, and \cite{kaufman2011} and \cite{gramacy2015} for applications to large datasets. The NNGP models defined here seamlessly extend to higher dimensional settings and may be attractive solutions to problems where a full GP is not feasible. This is illustrated using a simulated dataset comprising $n$=$2.5\times 10^4$ outcomes observed in a four-dimensional unit hypercube with an additional $5\times 10^3$ withheld for out-of-sample prediction. Outcomes for the $3\times 10^4$ randomly distributed locations were generated from $N(\bw\given \bzero, \bC + \taus \bI)$ where elements of $\bC$ were calculated using the Mat\'ern covariance function (\ref{eq:matern}), with $\sigma^2=1$, $\phi=10$, $\nu=1$, and $\tau^2=0.1$. The simulated data mean was set to zero to simplify the comparison described below.

Response and Conjugate model out-of-sample predictive performance is compared with that achieved using a local approximate Gaussian processes as implemented in the \texttt{laGP} R package \citep{lagp, gramacy2016laGP}. For the comparison, we used the \texttt{laGP} package routine \texttt{aGP} with default values for the Gaussian correlation function lengthscale parameter \texttt{d}, initial nugget \texttt{g}, and nearest neighbor selection objective function \texttt{method="mspe"}.

Both the Response and Conjugate models were fit using the Mat\'ern covariance function. Nearest neighbor ordering for both NNGP models was base on the sum of the four coordinate values. Parameter values for $\phi$, $\nu$, and $\alpha$ used in the Conjugate model were selected via five-fold cross-validation within the $2.5\times 10^4$ observations. Candidate parameter values were defined on a coarse grid of 100 combinations of $\phi \in [1, 5, 10, 15, 20]$, $\nu \in [0.5, 1, 1.5, 2]$, and $\alpha \in [0.01, 0.1, 0.5, 1, 1.5]$. 

Analysis results are presented in Table~\ref{exp3-results}. The Response model accurately estimated the parameter values used to generate the data. Similarly, the Conjugate model's minimum RMSPE and CRPS criteria (averaged over the five-fold cross-validation results) both selected the correct $\phi$, $\nu$, and $\alpha$. Predictive performance based on the $5\times 10^3$ holdout locations was comparable between the NNGP and \texttt{laGP} models, as reflected by near identical CRPS, RMSPE, and 95\% interval coverage and mean interval width (Table~\ref{exp3-results}). The 95\% interval coverage and interval width summaries for the Conjugate and \texttt{laGP} models were based on a 1.96$\times \widehat{\texttt{Var(y(\bs))}}$ margin of error. 

The three models were run using 12 CPUs. For \texttt{laGP} the number of CPUs was set via the \texttt{aGP} \texttt{omp.threads} argument. \texttt{laGP} delivered prediction for the holdout set in 3.5 minutes. The Conjugate model's five-fold cross-validation plus its final run using the optimal parameter set took 5.1 minutes. The Response model required 223 minutes to deliver $25\times10^3$ MCMC iterations. The large disparity in run time between this Response model and the timing results presented in Figure~\ref{n-timing} is due to the Mat\'ern covariance function's Gamma and Bessel functions. Although the Mat\'ern covariance function has attractive theoretical properties, it rarely (in our experience) yields improvements in inference---over simpler covariance functions in the Mat\'ern family---that warrant its use. Indeed, the Response model fit using and exponential covariance function to these data required 1/10th the run time and delivered nearly identical predictive performance, i.e., RMSPE=0.54, CRPS=0.31, 95\% PIC=92.69, and 95\% PIW=2.01.  

We also experimented with the number of folds, i.e., split-set cross-validations, needed to identify the optimal set of parameters for the Conjugate model. For the data considered here, the correct parameter set was identified using a single fold, which effectively drops the run time for the Conjugate model to 1.02 minutes.   

\begin{table}[!htbp]
\begin{center}
%% \begin{sidewaystable}[!htbp]
%% \centering
\caption {Simulated dataset, parameter credible intervals $50\%\, (2.5\%,97.5\%)$ and predictive validation.}\label{exp3-results}
\scriptsize
\begin{tabular}{ccccc}
\toprule
Parameter & True &  Response & Conjugate & \texttt{laGP}\\
\cmidrule{3-5}
$\sigma^2$  &1&0.97 (0.88, 1.08)&1&--\\ 
$\tau^2$  &0.1&0.10 (0.09, 0.12)&0.1&--\\
$\phi$  &10&11.05 (9.61, 12.55)&10&--\\
$\nu$  &1&1.10 (1.00, 1.22)&1&--\\ 
\midrule
CRPS  &&0.3&0.3&0.32\\
RMSPE  &&0.54&0.54&0.57\\
95\% PIC  &&92.78&94.66&93.32\\
95\% PIW  &&1.99&2.08&2.19\\
\bottomrule
\end{tabular} 
%% \end{sidewaystable}
\end{center}
\end{table}

We would like to point out that this application uses an isotropic model. However, most applications involving truly high-dimensional locations involve computer emulations where each parameter is considered a location co-ordinate. In such a case, anisotropic models seem to be more suitable, and then choosing the ``nearest neighbors'' becomes a conundrum as their is no unique (parameter-free) distance metric suitable for quantifying the separation between two such high-dimensional locations due to the scale mismatch among the coordinates of the location vector. A solution for the spatio-temporal case is offered in \cite{dnngp} where the anisotropic covariance function serves as a proxy for distance. However, this will introduce significant computational burden for the high-dimensional case. More research needs to be done to expedite anisotropic NNGP models for high-dimensional locations.

\end{document}